\documentclass[a4paper,fleqn,usenatbib]{mnras}

 \usepackage{newtxtext,newtxmath}

\usepackage[T1]{fontenc}

\usepackage{graphicx}	
\usepackage{amsmath}	
\usepackage{amssymb}	

\usepackage{xcolor}

\newcommand{\nproj}{1536} 

\title[Knowing the unknowns]{Knowing the unknowns: uncertainties in simple estimators of galactic dynamical masses}

\author[D.~J.~R.~Campbell et al.]{David~J.~R.~Campbell,$^{1}$\thanks{E-mail: d.j.r.campbell@durham.ac.uk}
Carlos~S.~Frenk,$^{1}$
Adrian~Jenkins,$^{1}$
Vincent~R.~Eke,$^{1}$
\newauthor
Julio~F.~Navarro,$^{2,3}$
Till~Sawala,$^{4}$
Matthieu~Schaller,$^{1}$
Azadeh~Fattahi,$^{2}$
Kyle~A.~Oman$^{2}$
\newauthor
and Tom~Theuns$^{1}$
\\
$^{1}$Institute for Computational Cosmology, Department of Physics, Durham University, South Road, Durham, DH1 3LE, UK \\
$^{2}$Department of Physics \& Astronomy, University of Victoria, Victoria, BC V8P 5C2, Canada \\
$^{3}$Senior CIfAR Fellow \\
$^{4}$Department of Physics, University of Helsinki, Gustaf H\"allstr\"omin katu 2a, FI-00014 Helsinki, Finland \\
}

\date{Accepted 2017 April 21. Received 2017 March 21; in original form 2016 March 14}

\pubyear{2017}

\begin{document}
\label{firstpage}
\pagerange{\pageref{firstpage}--\pageref{lastpage}}
\maketitle

\begin{abstract}
  The observed stellar kinematics of dispersion-supported galaxies are
  often used to measure dynamical masses.  Recently, several
  analytical relationships between the stellar line-of-sight velocity
  dispersion, the projected (2D) or deprojected (3D) half-light
  radius, and the total mass enclosed within the half-light radius,
  relying on the spherical Jeans equation, have been proposed. Here,
  we make use of the APOSTLE cosmological hydrodynamical simulations
  of the Local Group to test the validity and accuracy of such mass
  estimators for both dispersion and rotation-supported galaxies, for
  field and satellite galaxies, and for galaxies of varying masses,
  shapes, and velocity dispersion anisotropies. We find that the mass
  estimators of Walker~et~al.\ and Wolf~et~al.\ are able to recover
  the masses of dispersion-dominated systems with little systematic
  bias, but with a $1\sigma$ scatter of 25 and 23 percent,
  respectively. The error on the estimated mass is dominated by the
  impact of the 3D shape of the stellar mass distribution, which is
  difficult to constrain observationally. This intrinsic scatter
  becomes the dominant source of uncertainty in the masses estimated
  for galaxies like the dwarf spheroidal (dSph) satellites of the
  Milky Way, where the observational errors in their sizes and
  velocity dispersions are small. Such scatter may also affect the
  inner density slopes of dSphs derived from multiple stellar
  populations, relaxing the significance with which
  Navarro--Frenk--White profiles may be excluded, depending on the
  degree to which the relevant properties of the different stellar
  populations are correlated. Finally, we derive a new optimal mass
  estimator that removes the residual biases and achieves a
  statistically significant reduction in the scatter to 20 percent
  overall for dispersion-dominated galaxies, allowing more precise and
  accurate mass estimates.
\end{abstract}




\section{Introduction}
\label{sec:intro}

Obtaining reliable dynamical information lies at the heart of many
fundamental questions in cosmology and galactic structure. However, in
nature we typically only have partial information about the kinematics
of observed systems. Dynamical studies are therefore intrinsically
underconstrained, and often require the use of various simplifying
assumptions. In this study, we are interested in stellar dynamical
systems to which the spherical Jeans equation may be applied, such as
globular clusters, satellite galaxies, and massive spheroidal
galaxies. The use of Jeans analysis to infer the mass distribution of
a system whose gravitational potential is traced by stars is
complicated by an important degeneracy between the velocity dispersion
anisotropy of the stars, and the total mass profile of the system. The
anisotropy is notoriously difficult to constrain with current data,
and therefore studies based on the application of the spherical Jeans
equation have led to many ambiguous results. For example, a topic of
great interest is whether local satellite galaxies have central cusps
or cores in their central dark matter distributions. Studies based on
Jeans analysis are often inconclusive, largely due to the anisotropy
degeneracy.

An important advance in this subject has been the application of
simple estimators to infer the dynamical mass within a sphere of
radius equal to the projected (2D) stellar half-light radius,
$R_\mathrm{e}$ \citep{Walker_2009,Walker_2009_erratum}, or the
deprojected (3D) stellar half-light radius, $r_\mathrm{half}$
\citep{Wolf_2010}.\footnote{ $R_\mathrm{e}$ is the radius of the
  circle enclosing half of the total stellar luminosity in projection
  (effective radius), and $r_\mathrm{half}$ is the radius of the
  sphere enclosing the same luminosity fraction in 3D.}  These
estimators make use of only two measured quantities: the stellar
line-of-sight velocity dispersion averaged over the whole galaxy,
$\langle\sigma_\mathrm{los}\rangle$, and the projected half-light
radius. Each estimator can be written in the form,
\begin{equation}
  M(<\lambda R_\mathrm{e}) = \frac{\mu \langle\sigma_\mathrm{los}\rangle^2 R_\mathrm{e}}{G}~,
  \label{eq:est}
\end{equation}
where $M(<r)$ is the total mass enclosed within a sphere of radius
$r$. $\lambda$ and $\mu$ are dimensionless parameters, and $G$ is the
gravitational constant. These estimators are derived from the
spherical Jeans equation, which is valid for spherically symmetric,
dispersion-supported, collisionless, stationary systems, in dynamical
equilibrium, in the absence of streaming motions. Additional
assumptions are applied in the derivation of each estimator, with the
assumptions used by \citet{Walker_2009} being somewhat more
restrictive than those of \citet{Wolf_2010}.  The main finding of
these studies is that the total mass within a sphere of radius close
to the 2D or 3D stellar half-light radius is minimally sensitive to
the assumed form of the stellar velocity dispersion anisotropy and the
shape of the total mass profile (\citealt{Wolf_2010} discuss in detail
how this relatively tight constraint arises).

\citet{Walker_2011} have applied these estimators to apparently
distinct stellar subpopulations in the Fornax and Sculptor dwarf
spheroidals (dSphs), and have argued for the presence of a central
core in the dark matter density profile of these galaxies, with higher
confidence for Sculptor than for Fornax. (However, this result has
been disputed for Sculptor by \citealt{Strigari_2014}.) A further
example of the applicability of these mass estimators is the `too big
to fail problem' of the standard $\Lambda$ cold dark matter
($\Lambda$CDM) cosmological model \citep{BK_2011}, which was
predicated on the basis of comparing the circular velocity curves
($V_\mathrm{c}(r)=\sqrt{GM(<r)/r}$) of dark matter subhaloes drawn
from the Aquarius \citep{Springel_2008} and Via Lactea II simulations
\citep{Diemand_2008} with the dynamical mass within the 3D half-light
radius of the brightest satellites of the Milky Way (MW) according to
\citet{Wolf_2010}.

While the assumptions that underpin the spherical Jeans equation may
seem relatively benign, they are not exactly satisfied by all
dSphs. For example, it is clear that many galaxies, such as Fornax,
are not spherically symmetric. It also seems unlikely that satellites
orbiting in the potential well of a host halo are in dynamical
equilibrium.

The simple mass estimators have been tested on ideal spherically
symmetric systems \citep{Walker_2011}, simulations of ideal dSphs in a
static MW potential \citep{Kow_2013}, stellar distributions placed in
subhaloes from dark matter only simulations \citep{Laporte_2013_A},
and cosmological zoom simulations of isolated elliptical galaxies
\citep{Lyskova_2015}. More recently, the first generation of
simulations with enough resolution to model the satellite galaxies of
the MW reliably, in a realistic cosmological setting, have become
available \citep{Grand_2016,Sawala_2016,Wetzel_2016}. Such simulations
can be used to test the accuracy of equation~(\ref{eq:est}) in
estimating the dynamical masses of model galaxies with realistic
properties. In this paper, we make use of the APOSTLE simulations of
the Local Group \citep{Fattahi_2016,Sawala_2016} to study the accuracy
of the estimators proposed by \citet{Walker_2009} and
\citet{Wolf_2010}.

In Section~\ref{sec:est} we describe the mass estimators and the
assumptions on which they are based in greater detail. The simulations
used in this work are discussed in Section~\ref{sec:sims}.  In
Section~\ref{sec:general} we present general properties of a sample of
well-resolved dispersion-dominated galaxies drawn from the
simulations, to which we apply the mass estimators in
Section~\ref{sec:app}. The implications of the systematic errors on
the estimated masses are discussed in Section~\ref{sec:effects}. In
Section~\ref{sec:calib} we find the set of estimator parameters that
optimises the accuracy of the recovered mass for dispersion-dominated
galaxies. Concluding remarks are given in
Section~\ref{sec:conclusions}. In Appendix~\ref{sec:wolf3D} we
investigate the accuracy of the estimator of \citet{Wolf_2010} when
written in terms of the 3D half-light radius.  All simulation results
presented in this paper are at redshift zero. The assumed $\Lambda$CDM
cosmological parameters are given in Section~\ref{sec:apostle}.

\section{Mass estimators}
\label{sec:est}

The spherical Jeans equation relates the circular velocity curve of a
spherically symmetric system to the radial distribution and velocity
dispersion of a population of tracers that orbit in the gravitational
potential (see Section~\ref{sec:intro} for a list of the assumptions
on which this equation is based):
\begin{equation}
  \frac{G M(<r)}{r}
  = - \sigma_r^2(r) \left[
    \frac{\mathrm{d}\log\rho(r)}{\mathrm{d}\log r}
    + \frac{\mathrm{d}\log\sigma_r^2(r)}{\mathrm{d}\log r}
    + 2\beta(r)
  \right]~,
  \label{eq:jeans}
\end{equation}
where $M(<r)$ is the total mass enclosed within a radius $r$ of the
centre, $\rho(r)$ is the density profile of the tracer
population,\footnote{Strictly, $\rho(r)$ is the \textit{number}
  density profile of the tracer population, or indeed, the mass (or
  luminosity) density profile, provided that all tracers have the same
  mass (or luminosity).} $\sigma_r(r)$ is the profile of its velocity
dispersion, in the radial direction, and $\beta(r)$ is the tracer
velocity dispersion anisotropy, which encodes the balance between the
tangential and radial components of the velocity dispersion. At a
given radius,
\begin{equation}
  \beta(r) = 1 - \frac{\sigma_t^2(r)}{2\sigma_r^2(r)}~,
  \label{eq:beta}
\end{equation}
where $\sigma_t(r)$ is the tangential component of the velocity
dispersion, such that the total (3D) dispersion is given by,
\begin{equation}
  \sigma_\mathrm{3D}(r) =  \sqrt{\sigma_r^2(r) + \sigma_t^2(r)}~.
\end{equation}
Thus defined, $\sigma_t(r)$ includes the contributions to the total
velocity dispersion from two mutually orthogonal tangential
directions, which are equivalent under the assumption of spherical
symmetry.

When applied to a dSph galaxy, for example, equation~(\ref{eq:jeans})
relates the total mass profile to the spatial distribution and
kinematics of stars orbiting in the total gravitational potential
(which is dominated by the dark matter).  For an observed galaxy, we
can measure the projected stellar density, $\Sigma(R)$, and velocity
dispersion along the line of sight,\footnote{In this paper, a lower
  case $r$ is used to denote a 3D radius, and an upper case $R$ is
  used to denote a 2D, projected, radius.} $\sigma_\mathrm{los}(R)$.
A deprojection is required to map these observable profiles onto the
3D profiles that appear on the right-hand-side of
equation~(\ref{eq:jeans}). The projected and 3D stellar density
profiles, $\Sigma(R)$ and $\rho(r)$, under the assumption of spherical
symmetry, are related by an Abel transform,
\begin{equation}
  \rho(r) = - \frac{1}{\pi}\int_r^\infty \frac{\mathrm{d}\Sigma(R)}{\mathrm{d}R} \frac{\mathrm{d}R}{\sqrt{R^2-r^2}} ~.
  \label{eq:abel}
\end{equation}
Given only line-of-sight kinematic data for the stars, there exists an
inconvenient degeneracy between $\beta(r)$ and $\sigma_r(r)$, such
that \citep{BM_1982},
\begin{equation}
  \Sigma(R)\sigma_\mathrm{los}^2(R) = 2 \int_{R}^{\infty}\rho(r)\sigma_r^2(r)\left[1 - \frac{R^2}{r^2}\beta(r)\right] \frac{r\mathrm{d}r}{\sqrt{r^2 - R^2}} ~.
  \label{eq:sigmaproj}
\end{equation}
In this way, our ignorance of $\beta(r)$ influences both the
deprojection of $\sigma_\mathrm{los}(R)$ onto $\sigma_r(r)$ in
equation~(\ref{eq:sigmaproj}), and the subsequent inference of the
enclosed mass profile from equation~(\ref{eq:jeans}).

In order to make progress, some assumption about the stellar velocity
dispersion anisotropy is required, ranging from assuming isotropy
($\beta=0$), to exploring a wide range of plausible forms for
$\beta(r)$ in the fitting procedure (e.g.\ \citealt{Wolf_2010}).

Through Jeans analysis of eight of the brightest dSphs of the MW,
assuming constant $\beta$, \citet{Walker_2009,Walker_2009_erratum}
find that the total mass within a sphere of radius equal to the
projected stellar half-light radius (effective radius),
$R_\mathrm{e}$, is relatively well constrained (compared to smaller
and larger radii), and seemingly robust against the choice of $\beta$,
and of the assumed shape of the total density profile (see also
\citealt{Strigari_2007,Strigari_2008,Penarrubia_2008,Penarrubia_2008_erratum,Wolf_2010,Amorisco_2011}).
Assuming that the stars follow a Plummer density profile, with a
constant and isotropic velocity dispersion, \citet{Walker_2009}
propose a simple estimator for the total mass enclosed within a sphere
of radius $R_\mathrm{e}$, which arises immediately from
equation~(\ref{eq:jeans}) given these additional assumptions,
\begin{equation}
  M(<R_\mathrm{e}) = \frac{5 \langle\sigma_\mathrm{los}\rangle^2 R_\mathrm{e}}{2 G}~,
  \label{eq:walker}
\end{equation}
where $\langle\sigma_\mathrm{los}\rangle$ is the (assumed to be
constant) line-of-sight stellar velocity dispersion averaged over the
whole galaxy.  That is, a constraint on the dynamical mass (or
equivalently, a point on the circular velocity curve) can be obtained
using only the stellar half-light radius and a single value for the
stellar velocity dispersion, which does not require spatially resolved
kinematic data.

\citet{Wolf_2010}, on the other hand, argue that the 3D radius within
which the sensitivity of the enclosed mass to $\beta$ is minimized is
in fact $r_3$, the radius at which the logarithmic slope of the
stellar density profile, $\mathrm{d}\log\rho(r)/\mathrm{d}\log r$,
equals $-3$. Assuming that $r_3$ is approximately the stellar 3D
half-light radius, $r_\mathrm{half}$, these authors propose the
estimator,
\begin{equation}
  M(<r_\mathrm{half}) = \frac{3 \langle\sigma_\mathrm{los}\rangle^2 r_\mathrm{half}}{G}~,
  \label{eq:wolf3D}
\end{equation}
where they stress that $\langle\sigma_\mathrm{los}\rangle$ must be the
luminosity-weighted mean dispersion. This estimator is based on the
assumption that $\sigma_\mathrm{los}(R)$ remains relatively flat from
the centre of the system out to beyond $R_\mathrm{e}$, and that
$\beta(r)$ does not have an extremum within the stellar
distribution. To express equation~(\ref{eq:wolf3D}) entirely in terms
of observable quantities, \citet{Wolf_2010} make the further
simplifying assumption that $r_\mathrm{half} = 4 R_\mathrm{e} / 3$, in
which case,
\begin{equation}
  M\left(< \frac{4}{3} R_\mathrm{e}\right) = \frac{4 \langle\sigma_\mathrm{los}\rangle^2 R_\mathrm{e}}{G}~.
  \label{eq:wolf2D}
\end{equation}
This estimator is of the same form as equation~(\ref{eq:walker}), but
with $\lambda=4/3$ and $\mu=4$, compared to $\lambda=1$ and $\mu=2.5$,
in the notation of equation~(\ref{eq:est}).  It is worth noting that
if both equations~(\ref{eq:walker}) and~(\ref{eq:wolf2D}) apply, then
the enclosed dynamical mass increases by a factor of 1.6 from
$R_\mathrm{e}$ to $4R_\mathrm{e}/3$. We test the accuracy of
equations~(\ref{eq:walker}) and~(\ref{eq:wolf2D}) using simulated
galaxies in Section~\ref{sec:app}, and return to
equation~(\ref{eq:wolf3D}) in Appendix~\ref{sec:wolf3D}.

\section{Simulations}
\label{sec:sims}

We now describe the simulations used in this paper, followed by the
definitions employed for subhaloes and galaxies, and the resolution
limit we impose to ensure converged galaxy properties.

\subsection{APOSTLE simulations}
\label{sec:apostle}

The APOSTLE simulations use the `zoom' technique to resimulate 12
systems consistent with observational constraints on the properties of
the Local Group, as described in more detail by \citet{Fattahi_2016}
and \citet{Sawala_2016}. The regions were selected from a dark matter
only simulation of a cosmologically representative volume of comoving
side $100\,\mathrm{Mpc}$, using $\Lambda$CDM parameters consistent
with WMAP7 \citep{Komatsu_2011}. The density parameters at redshift
zero are $\Omega_\mathrm{m}=0.272$ (matter),
$\Omega_\mathrm{b}=0.0455$ (baryons), and
$\Omega_\Lambda=1-\Omega_\mathrm{m}=0.728$ (cosmological
constant). The present day Hubble parameter is
$H_0=100\,h\,\mathrm{km\,s^{-1}\,Mpc^{-1}}$, with $h=0.704$. The
linear power spectrum is normalised using $\sigma_8=0.81$ at redshift
zero. The spectral index of primordial fluctuations is
$n_\mathrm{s}=0.967$. The pairs of MW and M31 analogues were selected
on the basis of their separations, relative radial and tangential
velocities, and halo masses, along with the recession velocities of
outer Local Group members (see \citealt{Fattahi_2016} for details).

The 12 regions in the APOSTLE suite were simulated using the code from
the EAGLE project \citep{Crain_2015,Schaye_2015}, at a series of
resolution levels, which we label as LR, MR, and HR, in order of
increasing resolution (low, medium, high). Table~\ref{tab:sims} lists
the dark matter and (initial) gas particle masses in the zoom region
at each resolution level, along with the gravitational force
softening.  All regions have been simulated at LR and MR, and two
regions have also been run at HR. The EAGLE code is a version of the
smoothed particle hydrodynamics (SPH) code \textsc{gadget}
\citep{Springel_2005} that includes `subgrid' models for radiative gas
cooling and heating \citep{Wiersma_2009_A}, reionization, star
formation \citep{Schaye_2004,Schaye_2008}, stellar mass loss and metal
enrichment \citep{Wiersma_2009_B}, stellar feedback \citep{DV_2012},
black hole formation and mergers \citep{RG_2015}, and feedback from
active galactic nuclei \citep{Booth_2009}. See \citet{Crain_2015} and
\citet{Schaye_2015} for full details of the subgrid models.  The
hydrodynamics implementation used is \textsc{anarchy} (Dalla Vecchia
in preparation), which uses the conservative pressure-entropy SPH
formulation derived by \citet{Hopkins_2013}.  See
\citet{Schaller_2015} for a description of \textsc{anarchy} and the
impact of the changes with respect to the original \textsc{gadget} SPH
scheme on galaxy properties in the EAGLE simulations.  The APOSTLE
simulations use the `reference' EAGLE model parameters as described by
\citet{Schaye_2015}.

\begin{table}
  \centering
  \caption{Parameters for each resolution level in the APOSTLE
    simulations. $m_\mathrm{DM}$ and $m_\mathrm{gas}$ are the high
    resolution (zoom) dark matter and initial gas particle masses
    respectively. $\epsilon(z=0)$ is the Plummer-equivalent
    gravitational force softening at redshift zero. The gravitational
    force is Newtonian on scales larger than $2.8\epsilon$. There is a
    small amount of variation in the particle masses used in different
    simulations at a given resolution level; average values are quoted
    here (see \citealt{Fattahi_2016} for the individual values).}
  \label{tab:sims}
  \begin{tabular}{cccc}
    \hline
    Resolution
    &$m_\mathrm{DM}~[\mathrm{M}_\odot]$
    &$m_\mathrm{gas}~[\mathrm{M}_\odot]$
    &$\epsilon(z=0)~[\mathrm{pc}]$ \\
    \hline
    LR	&$7.1\times10^6$  &$1.4\times10^6$  &710 \\
    MR	&$5.8\times10^5$  &$1.2\times10^5$  &307 \\
    HR	&$3.7\times10^4$  &$7.4\times10^3$  &134 \\
    \hline
  \end{tabular}
\end{table}

\subsection{Halo finding and galaxy definition}
\label{sec:galaxy}

To identify haloes in the simulations, we first make use of the
friends-of-friends (FOF) algorithm, considering only dark matter
particles, with a linking length of 0.2 times the mean interparticle
separation \citep{Davis_1985}. Baryonic particles are assigned to the
same FOF group as their nearest dark matter particle. Each FOF group
is then processed using \textsc{subfind}, which identifies overdense
gravitationally self-bound `subhaloes'
\citep{Springel_2001,Dolag_2009}. The position of a subhalo is taken
to be that of the particle with the minimum value of the gravitational
potential. The `main subhalo' of a FOF group is that whose such
particle has the lowest potential in the group.\footnote{Note that we
  still refer to this main halo from \textsc{subfind} as a
  \textit{sub}halo, even though it constitutes the main component of
  the FOF group. For our purposes, the distinction between the main
  subhalo and other subhaloes in a FOF group is not important. It is
  typical in the literature to label the galaxy found in the main
  subhalo as the `central' galaxy of the FOF group, while other
  subhaloes host `satellite' galaxies. However, this nomenclature can
  be confusing in some cases, e.g.\ in the APOSTLE simulations the MW
  and M31 analogues can reside in the same FOF group.}  All other
subhaloes are embedded in the main subhalo (a given particle can
belong to at most one subhalo). As described by \citet{Schaye_2015},
we combine any two subhaloes whose separation is less than the stellar
half-mass radius of either subhalo, for separations of at most 3
physical kpc. This final adjustment is designed to absorb a very small
number of very low mass subhaloes that are dominated by a single
baryonic particle of unusually high mass (as a result of exceptional
stellar mass loss to a gas particle, or black hole growth).

We define a galaxy to be the set of subhalo star particles within a
spherical aperture of radius $r_\mathrm{gal}$ equal to 15 percent of
the virial radius, $r_{200}$, as measured from the subhalo
centre.\footnote{$r_{200}$ is the radius of the sphere that encloses a
  mean density equal to 200 times the critical density of the
  Universe.} For subhaloes where the actual value of $r_{200}$ is not
meaningful, e.g.\ for the subhalo of a satellite galaxy embedded in a
much larger halo, we adopt the value of $r_{200}$ obtained using a
relation between $r_{200}$ and the maximum value of the subhalo
circular velocity curve, $V_\mathrm{max}$, calibrated using main
subhaloes in the EAGLE Ref-L025N0752
simulation.\footnote{Ref-L025N0752 has the highest resolution
  available in the EAGLE simulation suite, which is similar to that of
  the MR APOSTLE simulations. The assumed cosmological parameters are
  slightly different to those used in the APOSTLE simulations, which
  is not important for our purposes.} In practice, we make use of the
directly measured value of $r_{200}$ when computing $r_\mathrm{gal}$
for all main subhaloes, and for any other subhalo for which $r_{200}$
is less than the distance to the furthest subhalo particle. All galaxy
properties presented in this paper are computed using the set of
subhalo star particles within $r_\mathrm{gal}$, with the galaxy centre
set to that of its host subhalo. We adopt the rest frame of the centre
of mass of the set of star particles defined in this way.

In each simulation volume, we shall refer to the MW and M31 analogues
as the `primary' galaxies. Galaxies within $300\,\mathrm{kpc}$ of the
centre of either primary galaxy are labelled as `satellites', and
those at larger distances are labelled as `field' galaxies. This
classification is independent of the particular FOF group in which a
galaxy resides.

\begin{figure}
  \centering
  \includegraphics[width=\columnwidth]{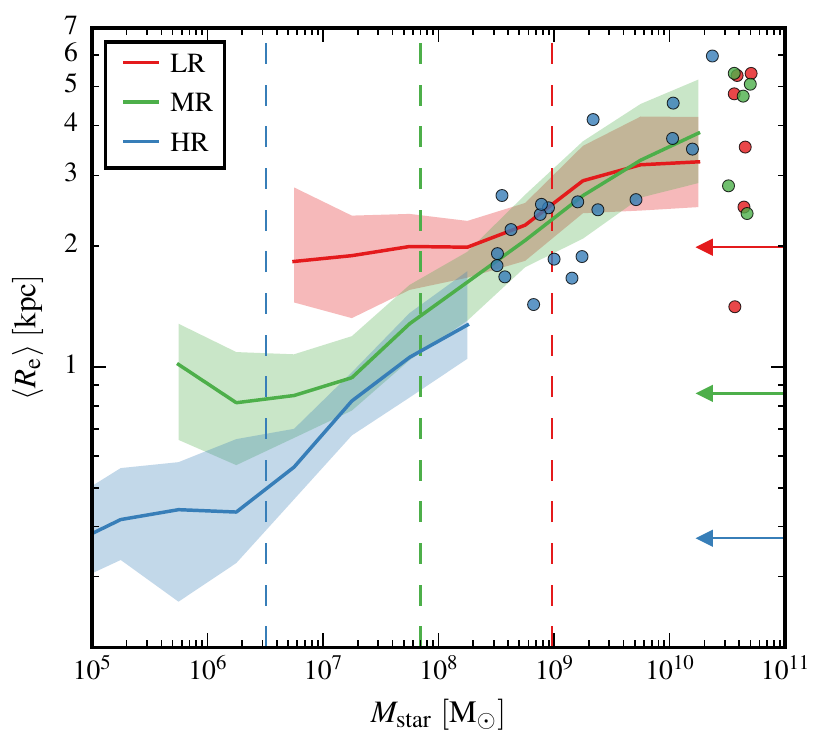}
  \caption{Mean projected stellar half-mass radius,
    $\langle R_\mathrm{e} \rangle$, versus stellar mass,
    $M_\mathrm{star}$, for all galaxies that have at least ten star
    particles in the APOSTLE simulations at each resolution level (12
    LR, 12 MR, and 2 HR simulations; different colours). The
    $\langle R_\mathrm{e} \rangle$ values are averaged over \nproj\
    evenly distributed projections. For each resolution, the median is
    shown as a solid line, and the $16^\text{th}-84^\text{th}$
    percentiles are indicated by a shaded region of the same
    colour. Points are shown instead for bins with fewer than ten
    galaxies. The vertical dashed lines show the minimum
    $M_\mathrm{star}$ of the set of galaxies composed of at least 1000
    star particles at each resolution. The horizontal arrows indicate
    2.8 times the gravitational softening at each resolution (see
    Table~\ref{tab:sims}).}
  \label{fig:rhalf_mstar_conv}
\end{figure}

\subsection{Selecting a well-resolved galaxy sample}
\label{sec:conv}

We assume that the luminosity of a star particle (which represents a
simple stellar population) is proportional to its mass. In this case,
$R_\mathrm{e}$ is the projected radius that encloses half the stellar
mass, as well as half the total stellar
luminosity. Fig.~\ref{fig:rhalf_mstar_conv} shows $R_\mathrm{e}$
versus stellar mass, $M_\mathrm{star}$, for all galaxies in the
APOSTLE simulations that are composed of at least ten star particles,
split by resolution level. The $R_\mathrm{e}$ values are averaged from
projecting over \nproj\ evenly distributed lines of sight for each
galaxy, making use of the HEALPix spherical tessellation
\citep{Gorski_2005}.\footnote{This corresponds to 3072 HEALPix
  pixels.} This set of sight-lines will be exploited throughout this
paper.

\begin{figure*}
  \centering
  \includegraphics[width=\textwidth]{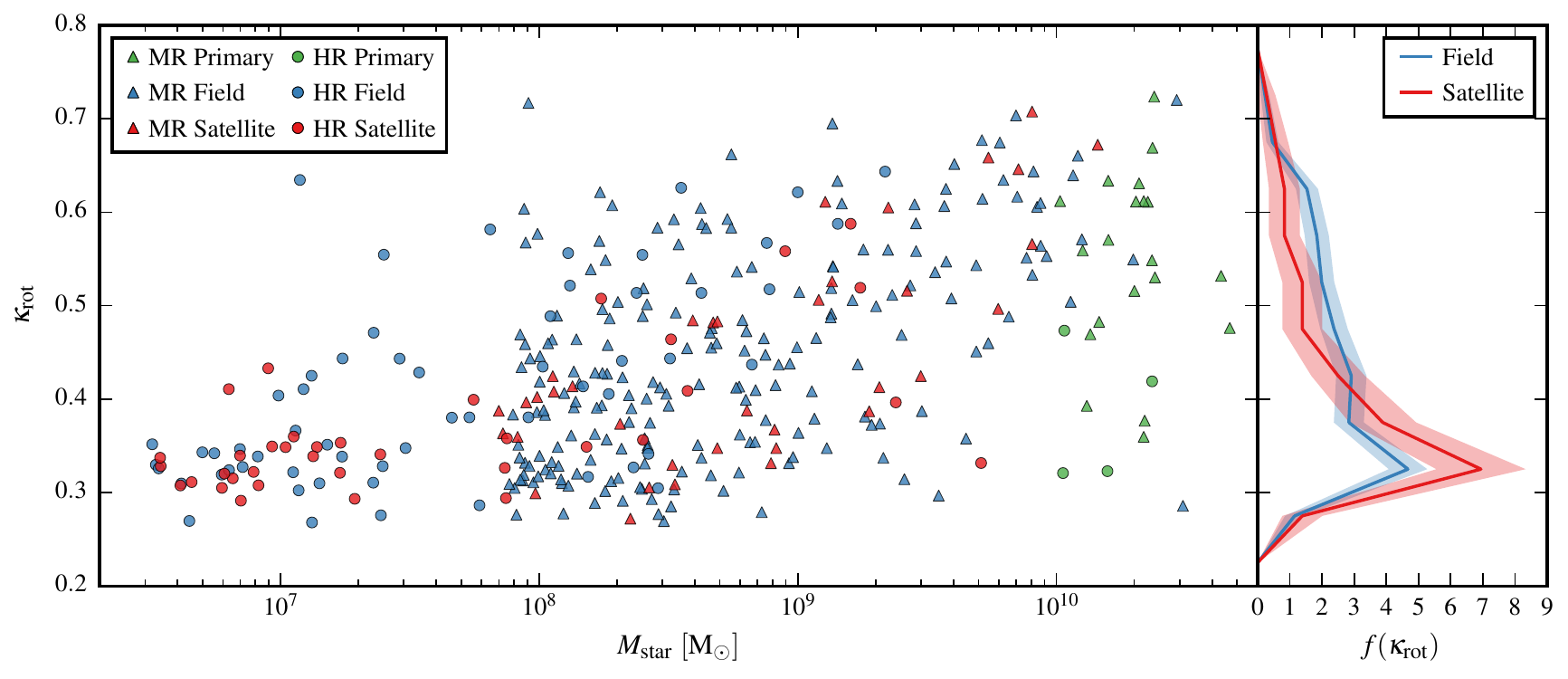}
  \caption{Kinematic measure, $\kappa_\mathrm{rot}$, used to
    discriminate amongst galaxies with different levels of stellar
    rotational support (see equation~\ref{eq:Krot}), shown versus
    stellar mass, $M_\mathrm{star}$, for all galaxies in our simulated
    sample. The galaxies are classified according to whether they are
    one of the MW or M31 analogues (primary), within
    $300\,\mathrm{kpc}$ of either of these (satellite), or located at
    larger distances (field).  The symbol shapes indicate the
    resolution level at which each galaxy has been simulated (MR or
    HR). Our dispersion-dominated galaxy sample consists of all
    galaxies that have $\kappa_\mathrm{rot}<0.5$. The panel on the
    right shows the distributions, $f(\kappa_\mathrm{rot})$, of
    $\kappa_\mathrm{rot}$ for all field and satellite galaxies. Each
    distribution is normalised to have unit area. The shaded regions
    show the $1\sigma$ error due to the Poisson noise on each bin
    count.}
  \label{fig:Krot_mstar}
\end{figure*}

In the following, we consider all galaxies in the highest resolution
realisation of each of the 12 APOSTLE regions (10 MR and 2 HR
simulations) that are resolved with at least 1000 star particles. This
conservative threshold has been chosen so that the stellar mass-size
relation is converged at each resolution level (see also
\citealt{Schaye_2015}). The vertical lines in
Fig.~\ref{fig:rhalf_mstar_conv} show the minimum $M_\mathrm{star}$ of
the set of galaxies with at least 1000 star particles at each
resolution. Note that the masses of star particles vary according to
the level of enrichment of the gas from which they formed, and the
extent of their own mass loss since their birth. There is also a small
amount of variation in the initial gas particle mass (and dark matter
particle mass) between different simulations at each resolution (see
Table~\ref{tab:sims}).

\begin{figure*}
  \centering
  \includegraphics[width=\textwidth]{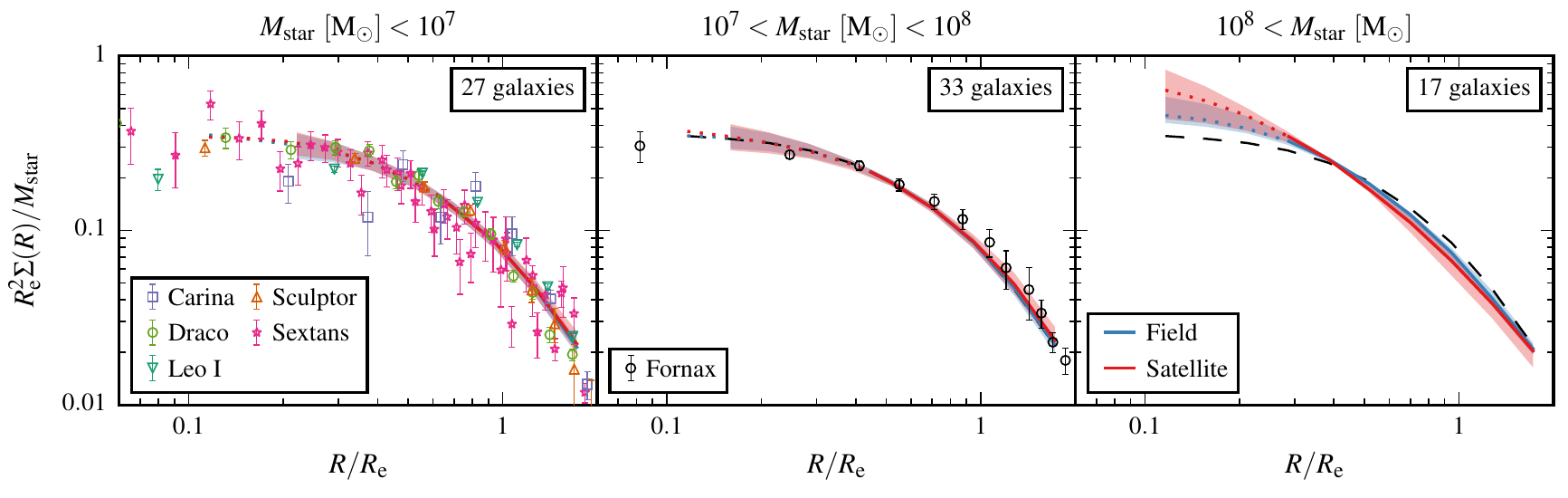}
  \caption{Projected stellar density profiles, $\Sigma(R)$, for
    dispersion-dominated ($\kappa_\mathrm{rot}<0.5$) field and
    satellite galaxies in the HR simulations. The solid lines are the
    median projected profiles obtained from \nproj\ evenly distributed
    projections for all galaxies of a given type (different colours;
    legend in right panel), around which the shaded regions of the
    same colour indicate the $16^\text{th}-84^\text{th}$ percentile
    spread. Each individual profile is scaled, prior to stacking, by
    the projected stellar half-mass radius, $R_\mathrm{e}$, for that
    line of sight, and the stellar mass, $M_\mathrm{star}$. The shaded
    regions are not shown below the gravitational softening (see
    Table~\ref{tab:sims}; in units of the median $R_\mathrm{e}$ in
    each panel), and the median profiles are dotted below 2.8 times
    the softening. The panels show different stellar mass ranges, and
    are labelled with the number of simulated galaxies they
    contain. As a visual aid, the median profile for all galaxies in
    the lowest mass bin is repeated as a dashed black line in the
    higher mass bins. The symbols with error bars show data for dSph
    satellites of the MW, where the data for each dSph are plotted in
    the relevant panel for its stellar mass following
    \citet{McConnachie_2012}. The data shown are for Carina
    \citep{Munoz_2006}, Draco \citep{Odenkirchen_2001}, Fornax
    \citep{Coleman_2005}, Leo~I \citep{Smolcic_2007}, Sculptor
    \citep{Battaglia_2008}, and Sextans \citep{Irwin_1995}. The
    measurements are scaled in the same way as the simulation
    predictions, assuming either a Gaussian (Leo~I) or Plummer (all
    others) density profile fit.}
  \label{fig:rho_pro}
\end{figure*}

\begin{figure*}
  \centering
  \includegraphics[width=\textwidth]{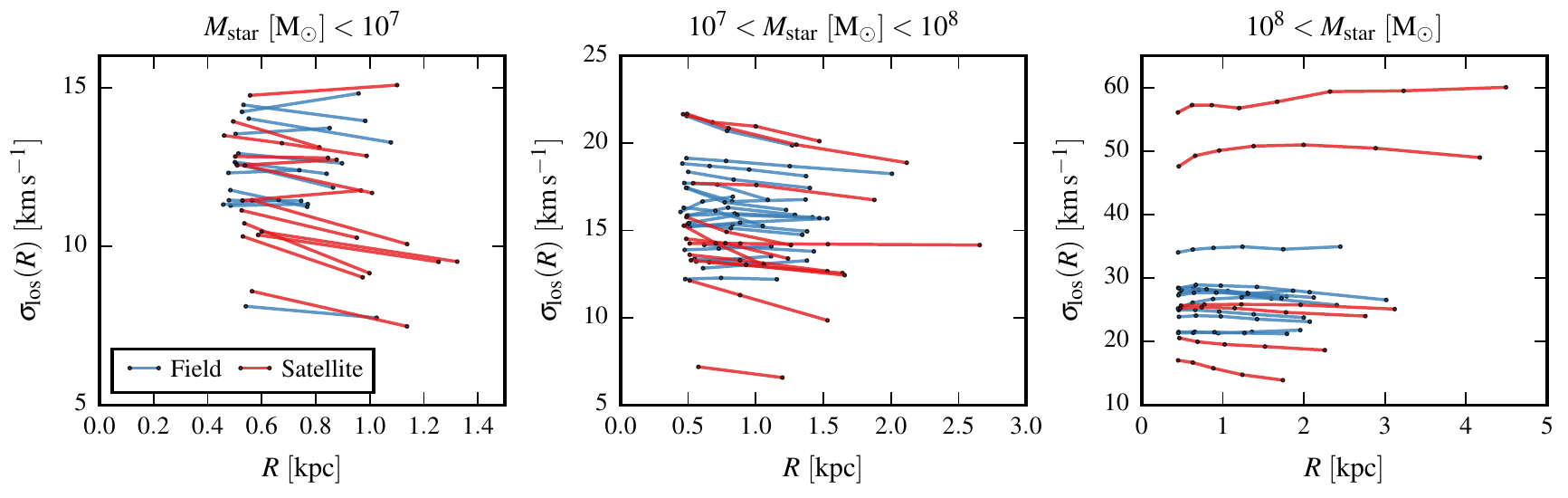}
  \caption{Line-of-sight stellar velocity dispersion profiles,
    $\sigma_\mathrm{los}(R)$, for dispersion-dominated
    ($\kappa_\mathrm{rot}<0.5$) field and satellite galaxies in the HR
    simulations, where each panel shows a different stellar mass
    range. The line shown for each galaxy is the median profile
    obtained from projecting over \nproj\ evenly distributed lines of
    sight, considering star particles at projected radii between 2.8
    times the gravitational softening (see Table~\ref{tab:sims}) and
    twice the mean projected stellar half-mass radius of the
    galaxy. The bin edges are evenly spaced in the logarithm of $R$,
    and the points shown are the linear means of the bin edges. The
    number of bins used for each galaxy is a function of its star
    particle count. Note that the axis limits are different for each
    panel.}
  \label{fig:sig_pro}
\end{figure*}

\begin{figure*}
  \centering
  \includegraphics[width=\textwidth]{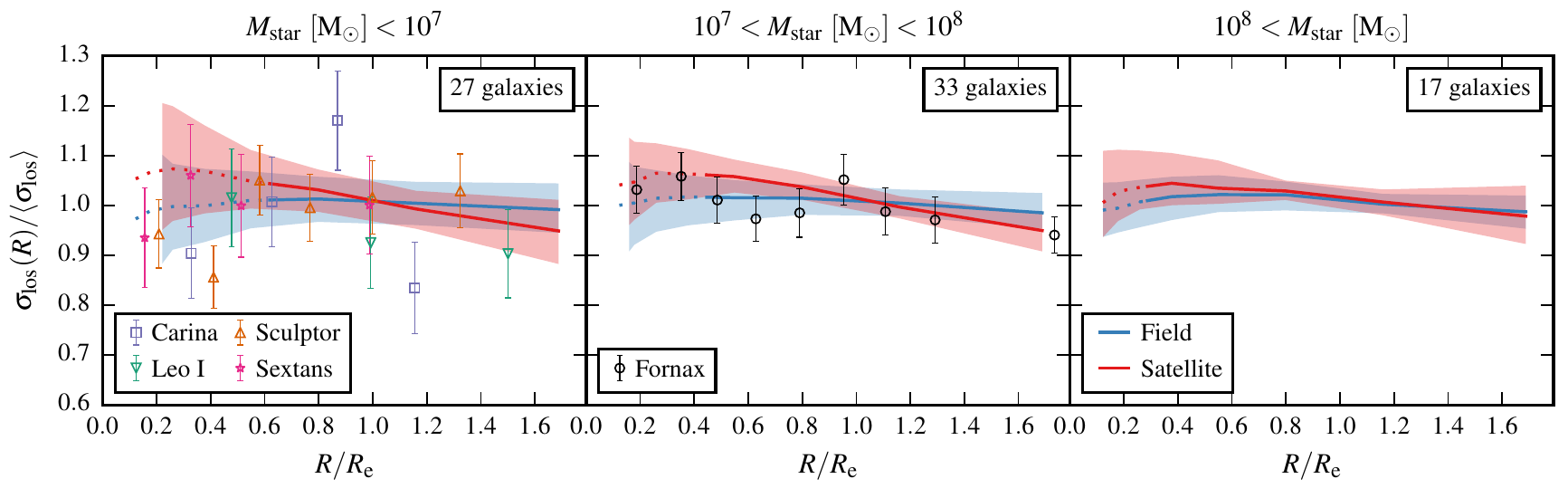}
  \caption{Line-of-sight stellar velocity dispersion profiles,
    $\sigma_\mathrm{los}(R)$, for dispersion-dominated
    ($\kappa_\mathrm{rot}<0.5$) field and satellite galaxies in the HR
    simulations, where each panel shows a different stellar mass range
    (labelled with the number of galaxies). The solid lines are the
    median profiles obtained from \nproj\ evenly distributed
    projections for all galaxies of a given type (different colours;
    legend in right panel), around which the shaded regions of the
    same colour indicate the $16^\text{th}-84^\text{th}$ percentile
    spread. Each individual profile is scaled, prior to stacking, by
    the projected stellar half-mass radius, $R_\mathrm{e}$, and the
    mean stellar velocity dispersion of the whole galaxy,
    $\langle\sigma_\mathrm{los}\rangle$, for that line of sight. The
    shaded regions are not shown below the gravitational softening
    (see Table~\ref{tab:sims}; in units of the median $R_\mathrm{e}$
    in each panel), and the median profiles are dotted below 2.8 times
    the softening. The symbols with error bars show data for dSph
    satellites of the MW, where the data for each dSph are plotted in
    the relevant panel for its stellar mass following
    \citet{McConnachie_2012}. The profiles shown are for Carina,
    Fornax, Leo~I, Sculptor, and Sextans, as computed by
    \citet{Strigari_2010}. The measurements are scaled in the same way
    as the simulation predictions, using the half-light radii from
    \citet{McConnachie_2012}.}
  \label{fig:sig_pro_stack}
\end{figure*}

\begin{figure*}
  \centering
  \includegraphics[width=\textwidth]{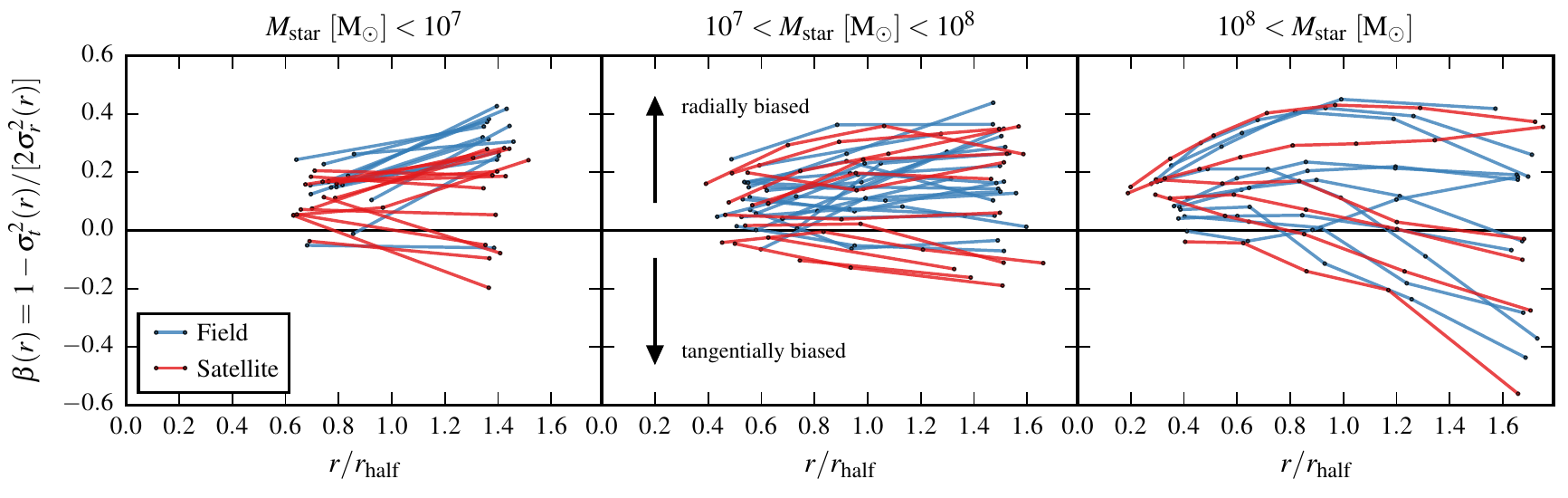}
  \caption{Spherically averaged stellar velocity dispersion anisotropy
    profiles, $\beta(r)$, for dispersion-dominated
    ($\kappa_\mathrm{rot}<0.5$) field and satellite galaxies in the HR
    simulations. $\beta(r)$ is negative for tangentially biased
    dispersions, zero for isotropy, and positive for radially biased
    dispersions (see equation~\ref{eq:beta}). The panels show
    different stellar mass ranges, as labelled. The profiles include
    all star particles between a radius of 2.8 times the softening
    (see Table~\ref{tab:sims}) and twice the 3D stellar half-mass
    radius, $r_\mathrm{half}$. The bin edges are evenly spaced
    percentiles of the radial distribution of the star particles, and
    the points show the median galactocentric radius of the star
    particles in each bin. The number of bins used for a given galaxy
    depends on how many star particles it has.}
  \label{fig:beta_pro}
\end{figure*}

\begin{figure*}
  \centering
  \includegraphics[width=\textwidth]{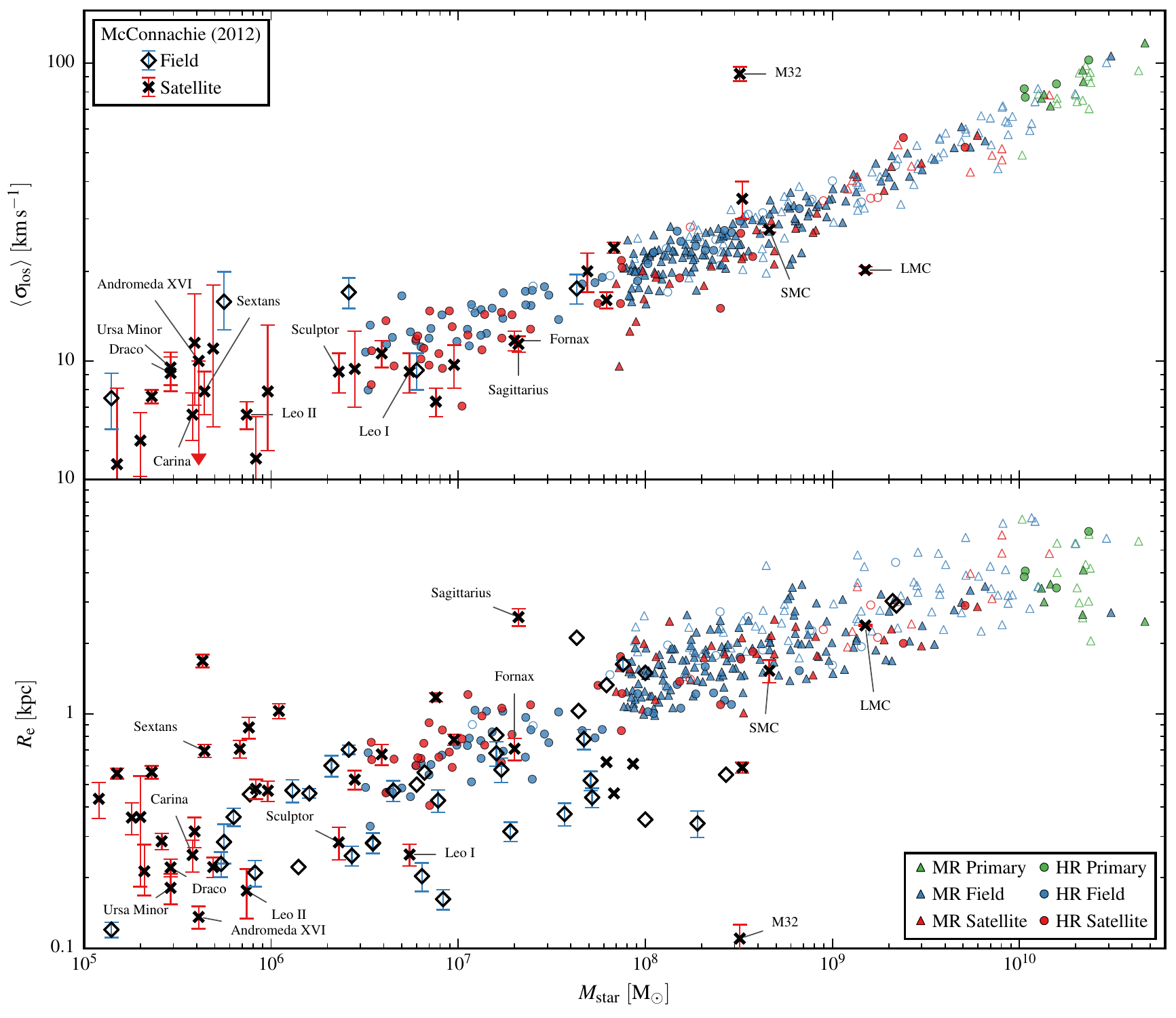}
  \caption{Line-of-sight stellar velocity dispersion,
    $\langle\sigma_\mathrm{los}\rangle$ (upper), and projected stellar
    half-mass radius, $R_\mathrm{e}$ (lower), versus stellar mass,
    $M_\mathrm{star}$, for all galaxies in our simulated sample.  The
    $\langle\sigma_\mathrm{los}\rangle$ and $R_\mathrm{e}$ values have
    been computed for a single, randomly chosen, line of sight through
    the simulations.  The galaxies are classified according to whether
    they are one of the MW or M31 analogues (primary), within
    $300\,\mathrm{kpc}$ of either of these (satellite), or at larger
    distances (field), as indicated in the lower right legend
    (different colours). The symbol shapes indicate the resolution
    level at which the galaxies have been simulated (MR or
    HR). Dispersion-dominated galaxies ($\kappa_\mathrm{rot}<0.5$) are
    plotted as filled symbols, while rotation-dominated galaxies
    ($\kappa_\mathrm{rot}>0.5$) are plotted as unfilled symbols of the
    same colour and shape. Data compiled by \citet{McConnachie_2012}
    for galaxies in this stellar mass range within $3\,\mathrm{Mpc}$
    of the Sun (excluding the MW and M31) are shown alongside the
    simulation predictions (upper left legend). The observed galaxies
    are categorised as satellites if they are associated with the MW
    or M31, otherwise they are classed as field galaxies. The error
    bars indicate the published uncertainties on the radii and
    velocity dispersions (where available). We supplement the
    \citet{McConnachie_2012} data with half-light radii for the SMC
    and LMC from \citet{Subramanian_2012} and \citet[disc model with
    no bar]{Weinberg_2001} respectively, integrating the published
    exponential density profile fits in each case. The labelled points
    are the 11 `classical' satellites of the MW, along with M32 (which
    has an exceptionally large dispersion and small size for its
    mass), and Andromeda~XVI (dispersion value is upper limit;
    downward arrow). We assume that the luminosity of a star particle
    is proportional to its mass, and therefore $R_\mathrm{e}$ is both
    the half-mass and half-light radius for the simulated galaxies.}
  \label{fig:rhalf_mstar_sig}
\end{figure*}

\begin{figure*}
  \centering
  \includegraphics[width=\textwidth]{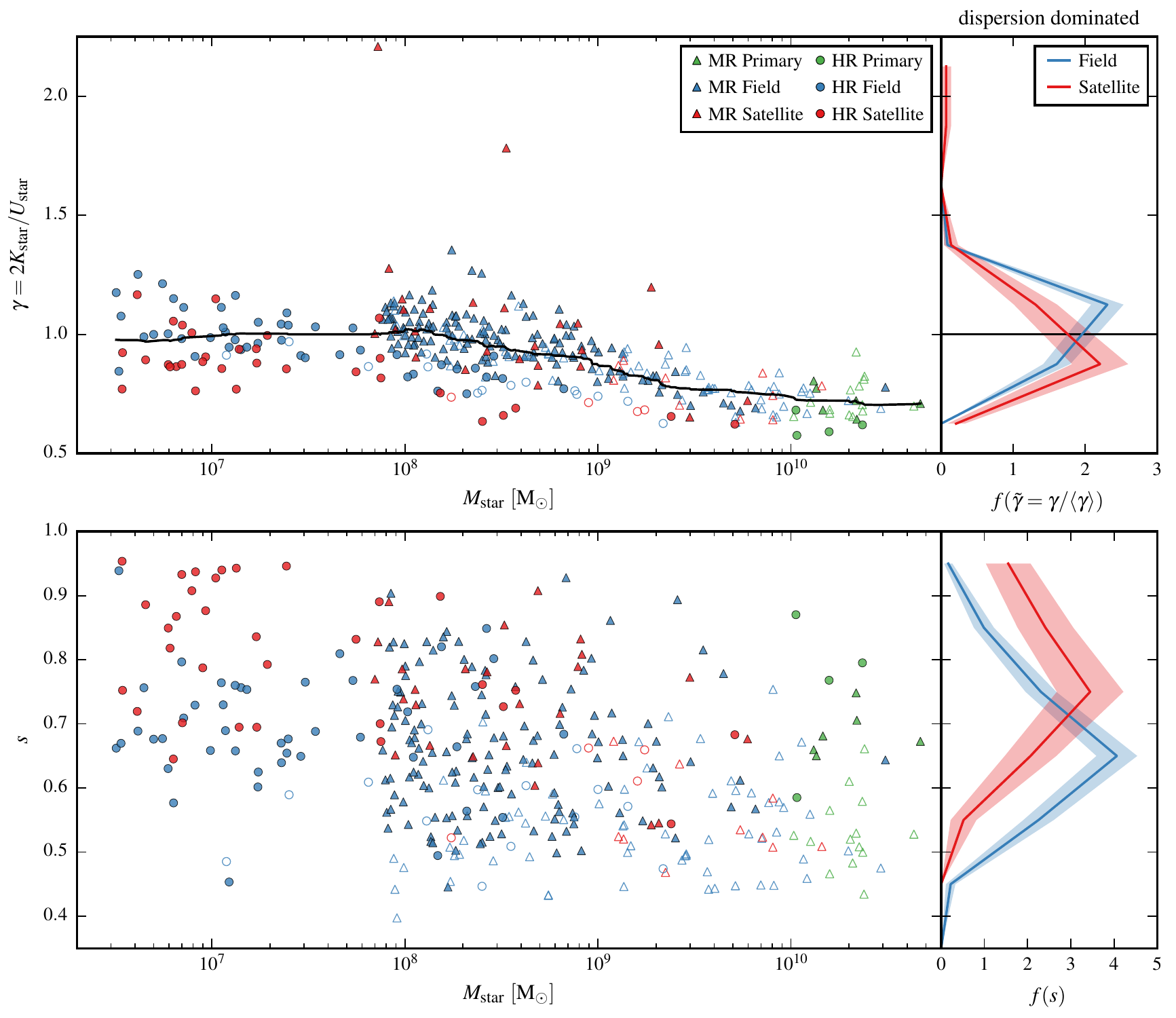}
  \caption{Stellar energy ratio, $\gamma$ (upper; see
    equation~\ref{eq:eqrat}), and stellar sphericity, $s$ (lower;
    derived from the reduced inertia tensor as defined in
    equation~\ref{eq:rit}), versus stellar mass, $M_\mathrm{star}$,
    for all galaxies in our simulated sample. The colours show whether
    a galaxy is one of the MW or M31 analogues (primary), within
    $300\,\mathrm{kpc}$ of either of these (satellite), or at larger
    distances (field). The symbol shapes indicate the resolution level
    (MR or HR). Dispersion-dominated galaxies
    ($\kappa_\mathrm{rot}<0.5$) are plotted as filled symbols, while
    the symbols are unfilled for rotation-dominated galaxies
    ($\kappa_\mathrm{rot}>0.5$). The black line in the upper panel
    shows the running median, $\langle\gamma\rangle$, of the $\gamma$
    values, considering at most 100 neighbouring galaxies in
    $M_\mathrm{star}$ for each galaxy. We define $\tilde{\gamma}$ as
    the ratio of $\gamma$ to this median line for each galaxy, and
    assume that galaxies with $\tilde{\gamma}\approx 1$ are in
    equilibrium. The panels on the right show the distributions of the
    rescaled energy ratio, $f(\tilde{\gamma})$ (\textit{not} the
    distribution of $\gamma$ itself), and the sphericity, $f(s)$, on
    the same vertical scales as the main panels, for only the
    dispersion-dominated field and satellite galaxies. Each
    distribution is normalised to have unit area. The shaded regions
    show the $1\sigma$ error due to the Poisson noise on each bin
    count.}
  \label{fig:mstar_spher_eqrat}
\end{figure*}

\section{General galaxy properties}
\label{sec:general}

In this section, we select a sample of simulated galaxies whose
stellar kinematics are dominated by dispersion. We then investigate
the basic properties of the galaxies in order to assess how realistic
they are. We present radial profiles of the projected stellar density
and line-of-sight velocity dispersion for galaxies that are resolved
at the highest resolution level, along with profiles of their stellar
velocity dispersion anisotropy. These profiles are central to the
mapping between observables and the total 3D mass profile (see
equations~\ref{eq:jeans}, \ref{eq:abel}, and~\ref{eq:sigmaproj}), and
thus represent the basic dynamical quantities that enter into the
Jeans analysis. We then place the simulated galaxies in the context of
the Local Group by comparing their stellar masses, projected half-mass
radii, and velocity dispersions with observational data. Finally, we
investigate the degree to which the simulated galaxies are spherically
symmetric and in dynamical equilibrium.

\subsection{Balance between dispersion and rotational support}
\label{sec:Krot}

A key assumption underpinning the spherical Jeans equation is that the
system under consideration is supported against gravitational collapse
by dispersion, rather than rotational or other streaming motion. This
clearly is not the case for galaxies with prominent stellar discs.  To
identify systems with significant rotational support, we make use of
the quantity $\kappa_\mathrm{rot}$ as introduced by \cite{Sales_2012},
which they define as `the fraction of kinetic energy invested in
ordered rotation'. $\kappa_\mathrm{rot}$ is computed as:
\begin{equation}
  \kappa_\mathrm{rot} = \frac{1}{K_\mathrm{star}}\sum_i\frac{m_i}{2}\left(\frac{j_{z,i}}{R_{xy,i}}\right)^2 ~,
  \label{eq:Krot}
\end{equation}
where $K_\mathrm{star}$ is the total stellar kinetic energy.  For each
star particle, $i$, of mass $m_i$, $j_{z,i}$ is the component of its
specific angular momentum in the direction of the total stellar
angular momentum vector, $\mathbfit{L}_\mathrm{star}$, and $R_{xy,i}$
is its distance from the axis ($z$) defined by
$\mathbfit{L}_\mathrm{star}$.

Strictly speaking, $\kappa_\mathrm{rot}$ is not directly sensitive to
ordered rotation, because the sign of the rotation about the $z$ axis
is lost in the squared term in equation~(\ref{eq:Krot}). Yet systems
that exhibit strong rotation have high values of
$\kappa_\mathrm{rot}$, which makes $\kappa_\mathrm{rot}$ a useful
measure to discriminate between systems that are dispersion-dominated
and those dominated by rotating discs. A pure disc galaxy with stars
on perfect circular orbits would have $\kappa_\mathrm{rot}=1$, while
instead $\kappa_\mathrm{rot}=1/3$ for a dispersion-supported system
with isotropic orbits (since $K_\mathrm{star}$ includes all three
orthogonal velocity components, but the summation in
equation~\ref{eq:Krot} considers only one such component). Thus,
$\kappa_\mathrm{rot}$ decreases from unity as random motion becomes
more important, with a lower limit in the region of
$\kappa_\mathrm{rot}\sim1/3$. However, dispersion-supported systems
with radially biased orbits can have $\kappa_\mathrm{rot}<1/3$, since
less of the kinetic energy is invested in tangential motion than in
the isotropic case.  Similarly, tangentially biased orbits imply
$\kappa_\mathrm{rot}>1/3$.

Fig.~\ref{fig:Krot_mstar} shows $\kappa_\mathrm{rot}$ versus
$M_\mathrm{star}$ for all galaxies in our sample (as defined in
Section~\ref{sec:conv}). It can be seen that the simulations predict a
broad range of stellar morphologies, according to this kinematic
measure. For $M_\mathrm{star} \gtrsim 10^9\,\mathrm{M}_\odot$,
rotation-dominated galaxies ($\kappa_\mathrm{rot}>0.5$) are slightly
more prevalent than those that are dominated by dispersion
($\kappa_\mathrm{rot}<0.5$). However, at progressively lower stellar
masses, the galaxies are progressively more likely to be
dispersion-dominated. For
$M_\mathrm{star} \lesssim 10^8\,\mathrm{M}_\odot$, only a small
fraction of galaxies have significant levels of rotational
support. The distributions of $\kappa_\mathrm{rot}$ for satellites and
field galaxies are not significantly different. The MR and HR
simulations predict similar distributions of $\kappa_\mathrm{rot}$ at
a given $M_\mathrm{star}$, over the common mass range explored.

In this paper, our main focus is on galaxies with dispersion-dominated
stellar kinematics, i.e.\ those with $\kappa_\mathrm{rot}<0.5$. The
galaxy sample defined in this way includes 70 percent of our full
sample.

\subsection{Stellar density and kinematic profiles}

Projected stellar density profiles, $\Sigma(R)$, for
dispersion-dominated field and satellite galaxies in the HR
simulations are shown in Fig.~\ref{fig:rho_pro}, split into bins of
stellar mass. The profiles are obtained by projecting over a large
number of lines of sight (as described in Section~\ref{sec:conv}), and
are stacked for clarity, showing field and satellite galaxies
separately. There is no significant difference between the profiles
for field and satellite galaxies in any mass bin. There is a small
variation in the shape of the profiles with increasing stellar mass,
such that the stellar density is more centrally concentrated in the
highest mass bin than in the other bins (see dashed line, which
repeats the median profile from the lowest mass bin). Projected
stellar number density profile data for bright dSphs of the MW are
shown alongside the simulation predictions. The measured profiles have
been rescaled assuming the best fitting of either a Plummer or
Gaussian density profile model (with $M_\mathrm{star}$ and
$R_\mathrm{e}$ as free parameters). We find that the data points for
Leo~I are best described by a Gaussian profile, while the other
galaxies shown each prefer a Plummer profile.  The simulation
predictions closely trace the observational data, in which the degree
of scatter varies for each dSph.  Thus, we can see that the simulated
galaxies have realistic stellar density distributions.

Line-of-sight velocity dispersion profiles, $\sigma_\mathrm{los}(R)$,
are shown in Fig.~\ref{fig:sig_pro} for the same set of simulated
galaxies.\footnote{When computing velocity dispersions, we weight by
  the particle mass. For example, the squared velocity dispersion in
  the $x$ direction is given by
  $\sigma_x^2 = (\sum_i m_i v_{x,i}^2) / (\sum_i m_i)$, where each
  particle, $i$, has mass $m_i$, and velocity $v_{x,i}$ in the $x$
  direction.} These profiles show each galaxy individually, using the
median result from all lines of sight. The profiles tend to be quite
flat with radius.  The typical dispersions and radial extents of the
galaxies scale closely with $M_\mathrm{star}$, and there are two
particularly large satellites with dispersions in excess of
$\sim50\,\mathrm{km\,s^{-1}}$ (note that there are two HR simulations,
and hence four primary galaxies, which between them host the
satellites shown). Recall that an assumption used in the derivation of
each of the mass estimators discussed in Section~\ref{sec:est} is that
$\sigma_\mathrm{los}(R)$ is relatively flat (or even constant). To see
the typical profile shapes more clearly, and to identify if there are
systematic differences in $\sigma_\mathrm{los}(R)$ for field and
satellite galaxies, we show scaled versions of these profiles in
Fig.~\ref{fig:sig_pro_stack}, averaging over all lines of sight in our
standard set. While the scatter in the profiles for both types of
galaxy is large, such that there is no significant difference between
the two populations, the median profiles are somewhat flatter for
field than for satellite galaxies, in the two smaller mass
intervals. The satellites in these panels tend to have relatively high
velocity dispersions in their centres, and relatively low dispersions
in their outer parts, compared with field galaxies. The median
profiles intersect near the projected half-mass radius.  The data
points in Fig.~\ref{fig:sig_pro_stack} show profiles for bright dSphs
of the MW \citep{Strigari_2010}. The simulated galaxies reproduce the
flatness in $\sigma_\mathrm{los}(R)$ observed for the dSphs.

Fig.~\ref{fig:beta_pro} shows spherically averaged profiles of the
velocity dispersion anisotropy, $\beta(r)$, defined in
equation~(\ref{eq:beta}), again examining the highest resolution
dispersion-dominated field and satellite galaxies, split by stellar
mass. The galaxies predominantly have radially biased dispersions
($\beta>0$), and only a small number of galaxies, which are in the
most massive subset, have $\beta\lesssim-0.2$ at any radius. The
profiles are similar in form for field galaxies and the satellites,
and there are no obvious distinctions between the two populations,
except that in the lowest mass bin the field galaxies tend to be
slightly more radially biased than the satellites, at both small and
large radii.  The anisotropy introduces a key degeneracy in Jeans
analysis (see Section~\ref{sec:est}), and the mass estimators of
interest here assume either that $\beta(r)$ is monotonic, or indeed
that $\beta$ has a negligible impact on the recovered mass, and so can
be assumed to equal zero.

\subsection{Galaxy sizes and integrated dispersions}

The dynamical mass estimators described in Section~\ref{sec:est} make
use of only two measurements: the projected stellar half-light radius,
and the line-of-sight stellar velocity dispersion. In
Fig.~\ref{fig:rhalf_mstar_sig} we plot these two quantities against
stellar mass for all galaxies in our sample. The simulations predict a
relatively tight relationship between
$\langle\sigma_\mathrm{los}\rangle$ and $M_\mathrm{star}$, while the
relationship between $R_\mathrm{e}$ and $M_\mathrm{star}$ exhibits a
somewhat higher level of scatter. There is clear convergence between
the results from the two resolution levels, which for the stellar
mass-size relation reflects our chosen threshold of 1000 star
particles (see Section~\ref{sec:conv}). There is no obvious difference
between the velocity dispersions or sizes of field and satellite
galaxies at a given stellar mass.

Observational data for galaxies within $3\,\mathrm{Mpc}$ of the Sun as
compiled by \citet{McConnachie_2012} are shown alongside the
simulation predictions in Fig.~\ref{fig:rhalf_mstar_sig}. Given our
conservative limit on the number of star particles required for a
galaxy to be included in our sample, we consider simulated galaxies
down to stellar masses close to that of Sculptor, although the
simulations do contain galaxies as faint as Draco. The simulation
predictions are consistent with the observational constraints, over
the large range in stellar mass shown.

It is apparent in Fig.~\ref{fig:rhalf_mstar_sig} that the simulations
predict somewhat less scatter in size at fixed stellar mass than seen
in the observational data. The majority of observed galaxies,
including the Magellanic Clouds and Fornax, lie on the predicted
relation but a substantial fraction are smaller, for their stellar
mass, than the smallest galaxies in the simulations. As shown in
Fig.~\ref{fig:rhalf_mstar_conv}, at each resolution level, the stellar
mass-size relation flattens out at low masses, and the minimum size
scales in the same way as the gravitational softening (the horizontal
arrows show the scale above which the force is Newtonian, at each
resolution). It thus appears that the resolution in the simulations is
not quite high enough to account for the sizes of all the observed
galaxies that have stellar masses close to those of the smallest
resolved galaxies in the simulations.  However, for our purpose, which
is to test the validity of dynamical mass estimators using the
simulated galaxies, it does not matter if resolution effects have had
a marginal impact on the galaxy sizes, provided that the galaxies have
self-consistent internal dynamics.

\subsection{Equilibrium and spherical symmetry}
\label{sec:eq_spher}

The spherical Jeans equation applies to spherically symmetric systems
that are in equilibrium. It is reasonable to expect that simple mass
estimators based on this equation will fail for galaxies that deviate
significantly from either of these assumptions.

In order to quantify the dynamical state of a galaxy in our sample, we
consider the ratio of the total kinetic and gravitational potential
energies of its star particles,
\begin{equation}
  \gamma = \frac{2 K_\mathrm{star}}{U_\mathrm{star}}~,
  \label{eq:eqrat}
\end{equation}
where $K_\mathrm{star}$ is the total stellar kinetic energy, and
$U_\mathrm{star}$ is the sum of the gravitational potential energy of
each star particle due to the full mass distribution of the subhalo to
which the star particles of the galaxy belong. In computing
$U_\mathrm{star}$, we shift the zero-point of the potential to
coincide with the galactic centre. Equation~(\ref{eq:eqrat}) is
similar in appearance to the virial ratio for an isolated system (such
that $\gamma$ would equal 1 if the stars were an isolated
self-gravitating system in equilibrium, with the potential taken to be
zero at an infinite distance from the system). However, $\gamma$
refers only to the stellar component, not to the whole dynamical
system, and we measure the potential energy with respect to the bottom
of the potential well. There is no natural $\gamma$ value for
equilibrium systems, and in the case of equilibrium the value of
$\gamma$ still depends on the form of the gravitational potential. The
upper panel of Fig.~\ref{fig:mstar_spher_eqrat} shows $\gamma$ versus
$M_\mathrm{star}$ for all galaxies in our sample. We consider a galaxy
to be out of equilibrium if it has a $\gamma$ value substantially
higher or lower than the median value for its stellar mass (black
line). We define $\tilde{\gamma}$ to be the ratio of $\gamma$ to the
median line shown for each galaxy. Note that the trend in $\gamma$
observed as a function of $M_\mathrm{star}$ in
Fig.~\ref{fig:mstar_spher_eqrat} does not imply a trend towards or
away from equilibrium.

In order to quantify how close to spherically symmetric a galaxy is,
we compute its stellar sphericity, $s$, as follows. We first compute
the reduced inertia tensor, $\mathbfss{I}$, of the star particles,
which projects the mass distribution onto a unit sphere, so that the
shape determination does not depend on the radial distribution of the
particles (e.g.\ \citealt{Bett_2012}). This symmetric tensor has
components $I_{ij}$, where $i,j\epsilon\lbrace 1,2,3 \rbrace$ such
that,
\begin{equation}
  I_{ij} = \frac{1}{M_\mathrm{star}}\sum_n m_n \frac{r_{n,i} r_{n,j}}{r_n^2} ~,
  \label{eq:rit}
\end{equation}
where each star particle, $n$, has mass $m_n$ and is located at a
distance $r_n$ from the galactic centre. $r_{n,i}$ and $r_{n,j}$ are
the coordinates of the star particle with respect to the galactic
centre, in the $i$ and $j$ directions, respectively. Taking the
eigenvectors of $\mathbfss{I}$ to be the principal axes of an
ellipsoid, with axis lengths $a \geq b \geq c$ given by the square
roots of the eigenvalues of $\mathbfss{I}$, the sphericity is $s=c/a$
(a sphere has $s=1$).  The lower panel of
Fig.~\ref{fig:mstar_spher_eqrat} shows $s$ versus stellar mass for all
galaxies in our sample. Naturally, the rotation-dominated galaxies
(unfilled symbols) tend be more flattened (and thus have lower $s$)
than those that are dispersion-dominated (filled symbols). However,
there are some galaxies which are dispersion-dominated according to
our kinematic measure ($\kappa_\mathrm{rot}$) that have lower
sphericity than the vast majority of rotation-dominated galaxies. Note
that $s$ does not distinguish between disc-like flattening and
elongation, as this measure takes into account only two of the axes.
The relationship between $s$ and $\kappa_\mathrm{rot}$ has a large
amount of scatter, although the scatter decreases (and $s$ decreases)
with increasing $\kappa_\mathrm{rot}$. Looking at
Fig.~\ref{fig:mstar_spher_eqrat}, it appears that satellites are
preferentially closer to spherical than field galaxies (see
distributions in right panel, which consider only the
dispersion-dominated galaxies).  This effect is consistent with the
expectation for satellite systems that have been tidally stripped
\citep{Barber_2015}.  The sphericity also increases on average as
stellar mass decreases in Fig.~\ref{fig:mstar_spher_eqrat}.

Given the broad range of 3D shapes and the spread in energy ratios
seen for the simulated galaxies in Fig.~\ref{fig:mstar_spher_eqrat},
there clearly exist galaxies within our (dispersion-dominated, or
indeed, full) sample for which the assumptions of spherical symmetry
and equilibrium are strongly violated. In the following, it will be
particularly interesting to see how the simple mass estimators perform
when applied to such galaxies.

\begin{figure}
  \centering
  \includegraphics[width=\columnwidth]{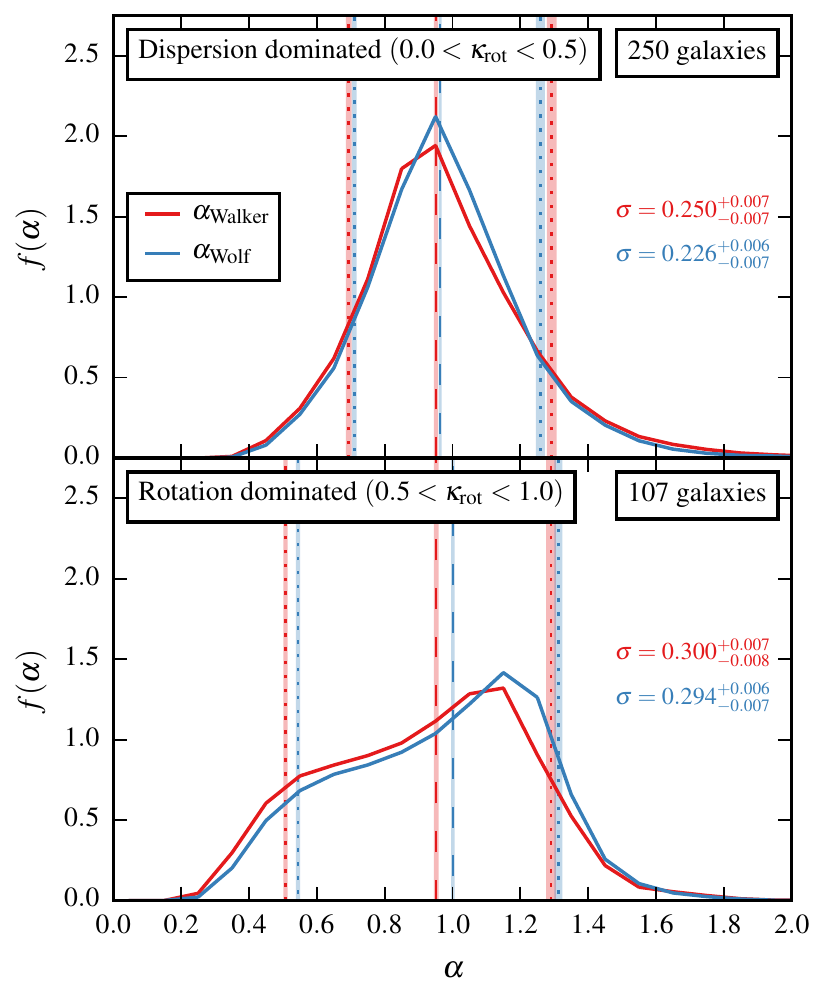}
  \caption{Distributions, $f(\alpha)$, of the estimated to true mass
    ratio, $\alpha$, for all galaxies, obtained by projecting over
    \nproj\ evenly distributed lines of sight. The galaxies are split
    into a dispersion-dominated sample (upper panel), and a
    rotation-dominated sample (lower panel), according to the value of
    $\kappa_\mathrm{rot}$ (see equation~\ref{eq:Krot}). Different
    colours show the distributions obtained using the
    \citet[$\alpha_\mathrm{Walker}$]{Walker_2009} and
    \citet[$\alpha_\mathrm{Wolf}$]{Wolf_2010} estimators, as
    labelled. The dashed lines indicate the median $\alpha$ values for
    each estimator and the dotted lines the $10^\text{th}$ and
    $90^\text{th}$ percentiles. The standard deviation, $\sigma$, of
    each distribution is given in the same colour as the lines.  The
    shaded regions around the vertical lines and the quoted errors on
    $\sigma$ are the $16^\text{th}-84^\mathrm{th}$ percentile
    confidence limits, derived from $10^4$ bootstrap samples of the
    galaxies for each distribution.  Each projection of each galaxy
    contributes to the relevant distribution with equal weight, and
    each distribution is normalised to have unit area. The panels are
    labelled with the number of galaxies they include. We highlight
    that the $f(\alpha)$ axis limits used in Figs.~\ref{fig:est_many},
    \ref{fig:est_many_halo}, \ref{fig:est_type},
    and~\ref{fig:calib_dist} vary with respect to those used here, in
    order to optimise the clarity of each individual figure.}
  \label{fig:est_Krot}
\end{figure}

\section{Accuracy of mass estimators}
\label{sec:app}

Having demonstrated that the galaxies in our simulations have
realistic projected stellar density and velocity dispersion
distributions, with a range of 3D shapes, velocity dispersion
anisotropies, levels of dispersion support, and departures from
dynamical equilibrium, we now apply the simple mass estimators
proposed by \citet{Walker_2009} and \citet{Wolf_2010} in order to
assess their accuracy as a function of various galaxy properties.

For convenience, let us denote the ratio of the estimated mass to the
true dynamical mass within some sphere as $\alpha$. Specifically, for
the estimator of \citet{Walker_2009}, from equation~(\ref{eq:walker}),
\begin{equation}
  \alpha_\mathrm{Walker} = \frac{5 \langle\sigma_\mathrm{los}\rangle^2 R_\mathrm{e}}{2 G M(<R_\mathrm{e})} ~,
  \label{eq:alpha_walker}
\end{equation}
and, for the estimator of \citet{Wolf_2010}, from
equation~(\ref{eq:wolf2D}),
\begin{equation}
  \alpha_\mathrm{Wolf} = \frac{4 \langle\sigma_\mathrm{los}\rangle^2 R_\mathrm{e}}{G M(<4R_\mathrm{e}/3)} ~.
  \label{eq:alpha_wolf}
\end{equation}
We emphasize that in both equations~(\ref{eq:alpha_walker})
and~(\ref{eq:alpha_wolf}), the denominator is the total mass within a
sphere of radius proportional to the value of $R_\mathrm{e}$ obtained
for a given line of sight. For each galaxy in our sample, we measure
$R_\mathrm{e}$, $\langle\sigma_\mathrm{los}\rangle$,
$M(<R_\mathrm{e})$, and $M(<4R_\mathrm{e}/3)$ for \nproj\ evenly
distributed lines of sight. The masses are obtained from all subhalo
particles within the relevant radius, including the contributions from
dark matter, gas, stars, and black holes, where present.

\subsection{Dispersion support}

The mass estimators assume that the dynamical system is supported by
dispersion.  We begin by applying the estimators to all galaxies in
our sample, including those with prominent stellar discs (for which we
would not expect an observer to use such an estimator).
Fig.~\ref{fig:est_Krot} shows the distributions of $\alpha$ values
obtained by applying both estimators to all projections of every
galaxy, split according to whether the stellar motions within the
galaxy are dominated by dispersion or rotation (see
Section~\ref{sec:Krot}).

For both kinematic regimes, each estimator has a bias in the median of
no more than 5 percent (dashed lines), accompanied by a large scatter
(see dotted lines and standard deviation values). It is interesting
that there is no dramatic change in the accuracy of the estimators, in
the median, when switching from the dispersion-dominated to the
rotation-dominated galaxies. For each sample, the two different
estimators have a similar level of scatter, which is larger for the
rotation-dominated galaxies. In general, the distributions are very
similar in shape for the two estimators; however, the
\citet{Wolf_2010} estimator has a smaller median offset,
$10^\text{th}$ to $90^\text{th}$ percentile spread, and standard
deviation than the \citet{Walker_2009} estimator, for both galaxy
samples (although the difference in the scatter is much less
significant for the rotation-dominated galaxies). The $\alpha$
distributions for the dispersion-dominated galaxies are approximately
symmetric around the median, but the shape of the distributions is
more complicated in the rotation-dominated case, where the peaks are
above $\alpha=1$, and there are extended tails to low $\alpha$.

Overall, there is a $1\sigma$ scatter for dispersion-dominated
galaxies of 25 percent when using the estimator of
\citet{Walker_2009}, and 23 percent for the estimator of
\citet{Wolf_2010}. In the following, we shall investigate the
dependence of the $\alpha$ distributions for dispersion-dominated
galaxies on key quantities that characterise the properties of each
galaxy and are most relevant to the assumptions on which the mass
estimators are based.

\begin{figure*}
  \centering
  \includegraphics[width=\textwidth]{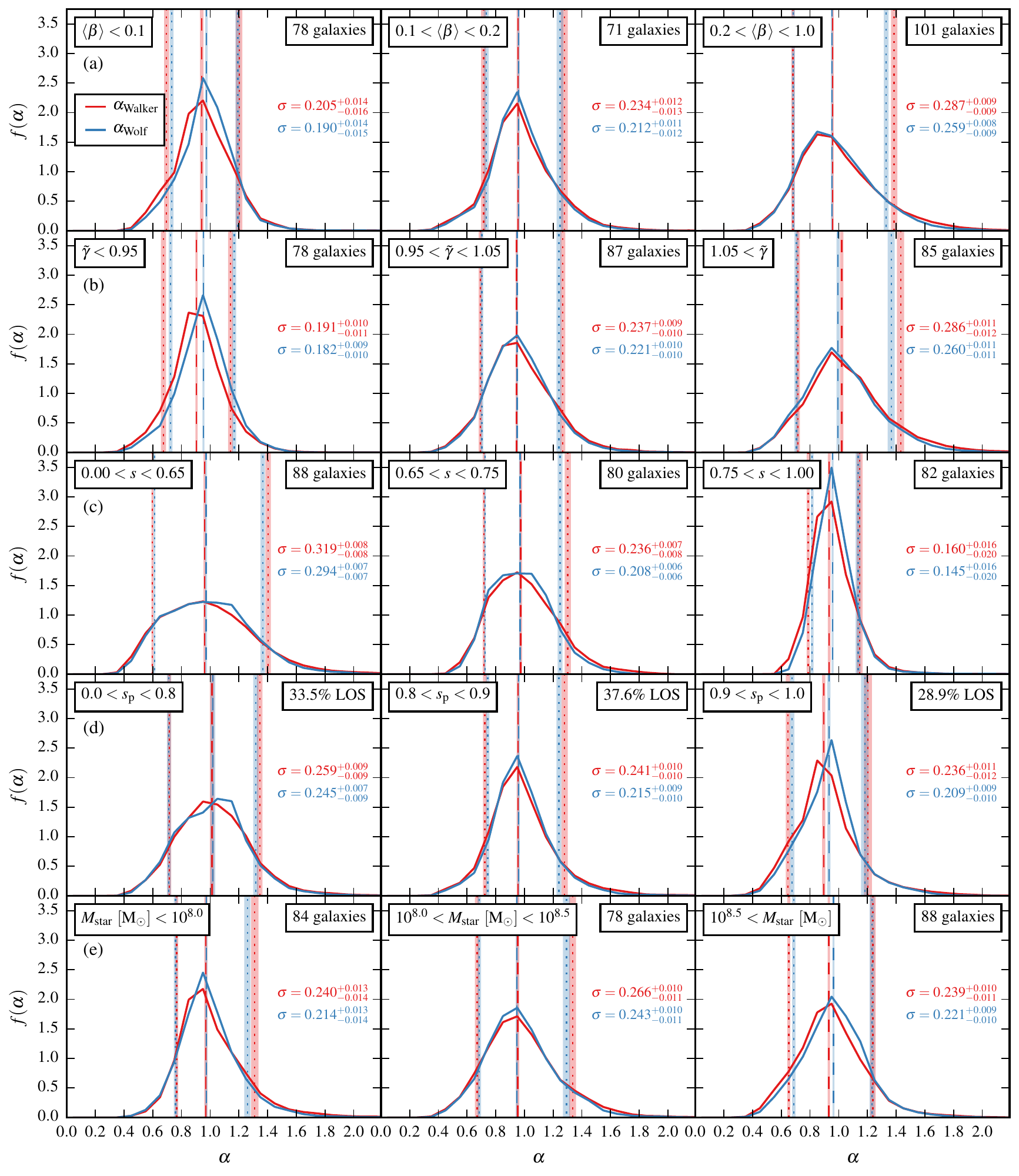}
  \caption{Distribution, $f(\alpha)$, of the estimated to true mass
    ratio, $\alpha$, for dispersion-dominated galaxies
    ($\kappa_\mathrm{rot}<0.5$), as a function of various galaxy
    properties. In each row, the dispersion-dominated sample is split
    into three bins (columns), according to: (a) the mean stellar
    velocity dispersion anisotropy, $\langle\beta\rangle$; (b) the
    stellar equilibrium measure, $\tilde{\gamma}$; (c) the stellar
    sphericity, $s$; (d) the projected stellar circularity,
    $s_\mathrm{p}$; and (e) the stellar mass, $M_\mathrm{star}$. Each
    panel is labelled with the interval it considers and the number of
    galaxies it contains (percentage of lines of sight in the case of
    circularity, where a single galaxy may contribute to more than one
    panel). The details of the computation of the distributions and
    the meaning of each line are as described in the caption of
    Fig.~\ref{fig:est_Krot}.}
  \label{fig:est_many}
\end{figure*}

\begin{figure*}
  \centering
  \includegraphics[width=\textwidth]{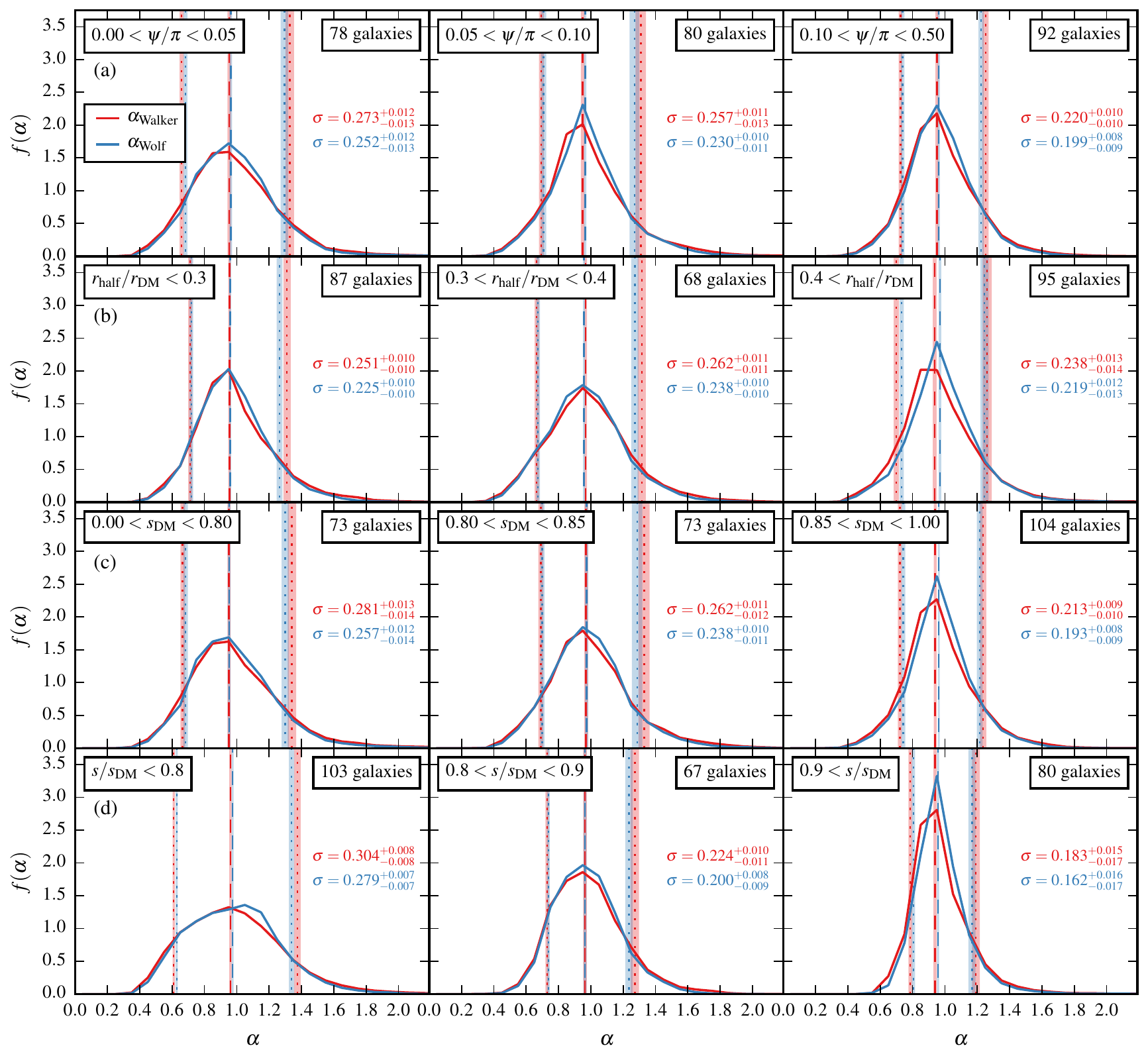}
  \caption{Distribution, $f(\alpha)$, of the estimated to true mass
    ratio, $\alpha$, for dispersion-dominated galaxies
    ($\kappa_\mathrm{rot}<0.5$), as a function of various galaxy and
    host subhalo dark matter properties. In each row, the
    dispersion-dominated galaxy sample is split into three bins
    (columns), according to: (a) the angle between the minor axis of
    the galaxy and that of the dark matter mass distribution within
    its host subhalo, $\psi$; (b) the ratio of the stellar 3D
    half-mass radius, $r_\mathrm{half}$, to the subhalo dark matter
    scale radius, $r_\mathrm{DM}$; (c) the subhalo dark matter
    sphericity, $s_\mathrm{DM}$; and (d) the ratio $s/s_\mathrm{DM}$,
    where $s$ is the stellar sphericity. Each panel is labelled with
    the interval it considers and the number of galaxies it
    contains. The details of the computation of the distributions and
    the meaning of each line are as described in the caption of
    Fig.~\ref{fig:est_Krot}.}
  \label{fig:est_many_halo}
\end{figure*}

\begin{figure}
  \centering
  \includegraphics[width=\columnwidth]{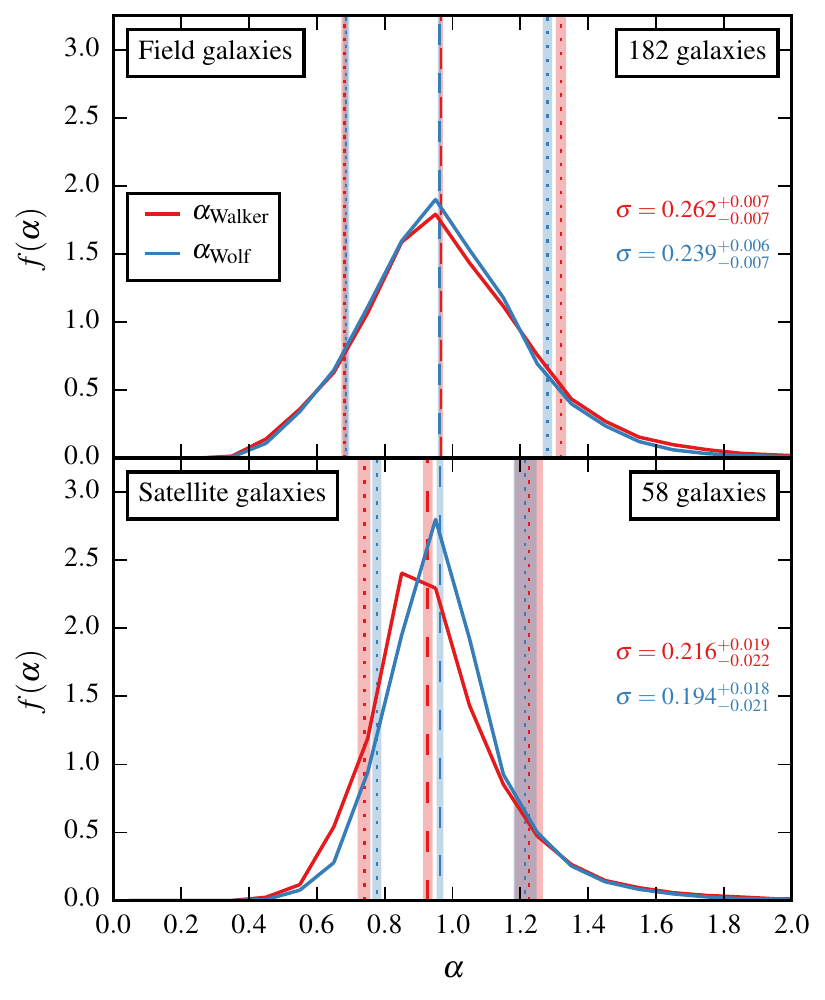}
  \caption{Distributions, $f(\alpha)$, of the estimated to true mass
    ratio, $\alpha$, for field (upper panel) and satellite (within
    $300\,\mathrm{kpc}$ of MW or M31; lower panel) galaxies in the
    dispersion-dominated sample ($\kappa_\mathrm{rot}<0.5$). The
    details of the computation of the distributions and the meaning of
    each line are as described in the caption of
    Fig.~\ref{fig:est_Krot}.}
  \label{fig:est_type}
\end{figure}

\begin{figure*}
  \centering
  \includegraphics[width=\textwidth]{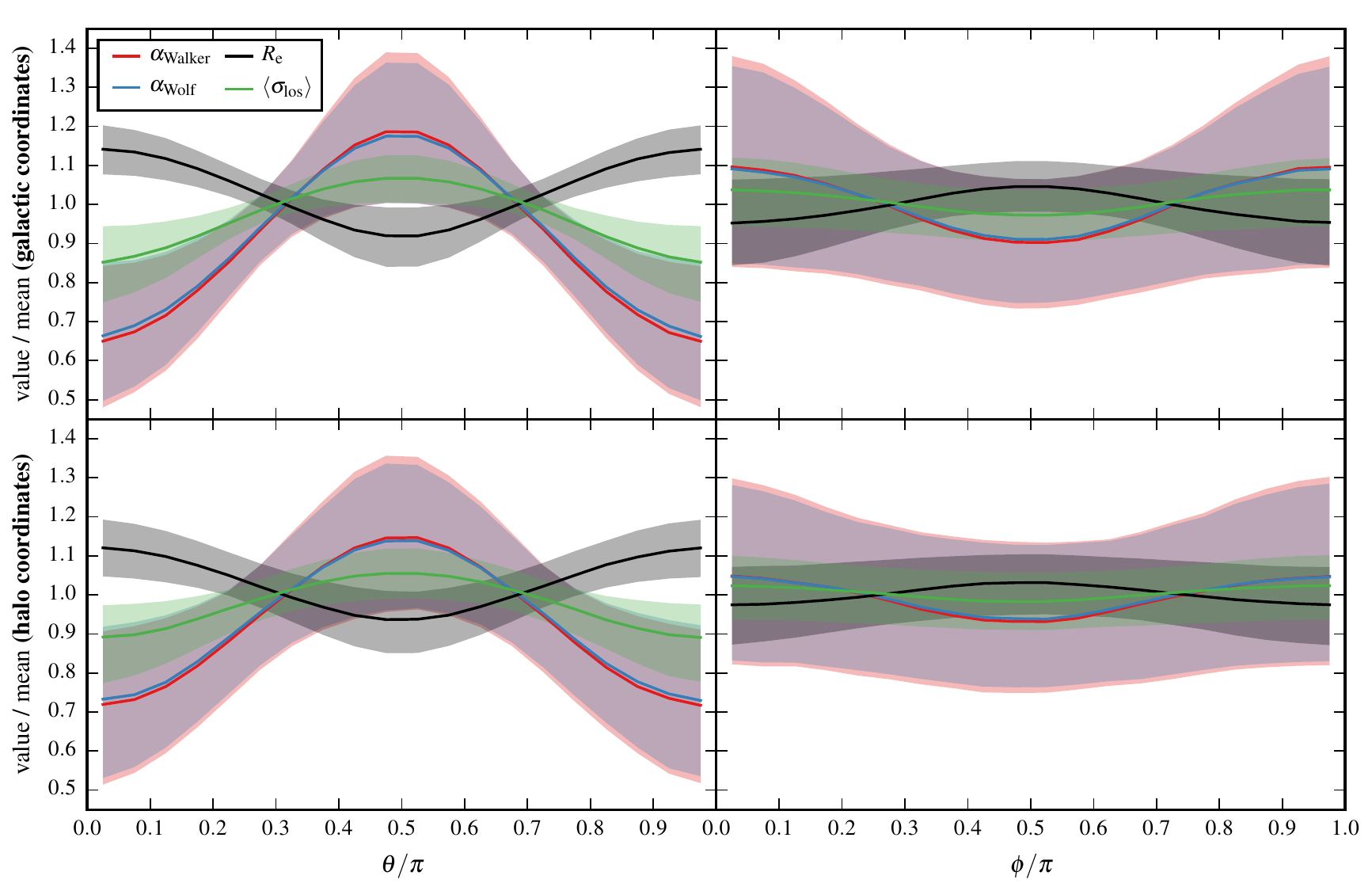}
  \caption{Angular variation of the estimated to true mass ratio for
    the \citet[$\alpha_\mathrm{Walker}$]{Walker_2009} and
    \citet[$\alpha_\mathrm{Wolf}$]{Wolf_2010} estimators, the
    projected stellar half-mass radius, $R_\mathrm{e}$, and the
    line-of-sight stellar velocity dispersion,
    $\langle\sigma_\mathrm{los}\rangle$, for dispersion-dominated
    galaxies ($\kappa_\mathrm{rot}<0.5$), computed by projecting over
    \nproj\ evenly distributed lines of sight. The curves are shown as
    functions of the spherical polar angles, $\theta$ and $\phi$,
    where the coordinate system is aligned with the eigenbasis of the
    reduced inertia tensor of either the galactic star particles
    (upper panels) or the dark matter particles of the host subhalo
    (lower panels). The $z$ axis ($\theta=0$) is aligned with the
    shortest principal axis, and the $x$ axis ($\theta=\pi/2,\phi=0$)
    is aligned with the longest principal axis. The spherical polar
    angular ranges are defined as $0\leq\theta\leq\pi$ and
    $0\leq\phi<2\pi$. Since projections in opposite directions are
    equivalent in this paper, the \nproj\ unique lines of sight are
    identified by their angular coordinates in the half-sphere defined
    by $0\leq\theta<\pi$ and $0\leq\phi<\pi$ within this figure. For
    each galaxy, the values of $\alpha_\mathrm{Walker}$,
    $\alpha_\mathrm{Wolf}$, $R_\mathrm{e}$, and
    $\langle\sigma_\mathrm{los}\rangle$ for each projection have been
    divided by the corresponding mean value over all projections of
    that galaxy, before computing the curves shown. The solid lines
    show the median values, and the shaded regions of the same colour
    indicate the $16^\text{th}-84^\text{th}$ percentile spread (the
    $\theta$ and $\phi$ axes use the same fixed bin width, in units of
    $\pi$). Each panel makes use of the full set of \nproj\ lines of
    sight, so at fixed $\theta$ (left) or $\phi$ (right), the median
    and percentile range values shown result from the data for all
    lines of sight (with that particular $\theta$ or $\phi$) for all
    galaxies.}
  \label{fig:est_angle}
\end{figure*}

\begin{figure}
  \centering
  \includegraphics[width=\columnwidth]{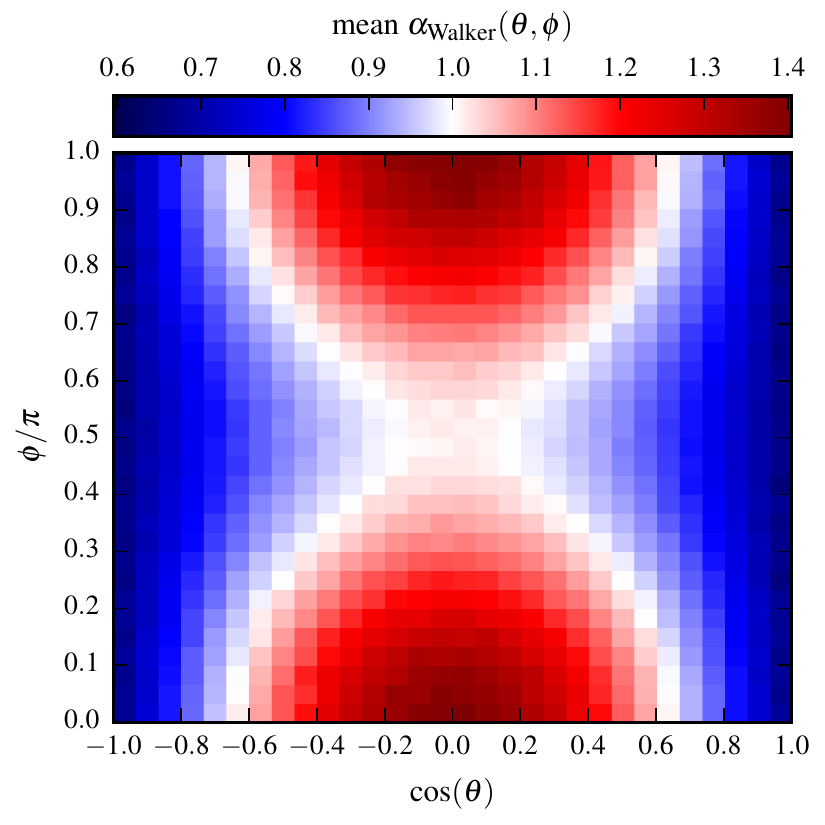}
  \caption{Estimated to true mass ratio,
    $\alpha_\mathrm{Walker}(\theta,\phi)$, resulting from applying the
    estimator of \citet{Walker_2009} to all galaxies in our
    dispersion-dominated sample ($\kappa_\mathrm{rot}<0.5$), by
    projecting over \nproj\ evenly distributed lines of sight for each
    galaxy. $\theta$ and $\phi$ are the spherical polar angles in the
    galactic coordinate system defined by the eigenbasis of the
    stellar reduced inertia tensor, where the $z$ axis ($\theta=0$) is
    aligned with the shortest stellar principal axis, and the $x$ axis
    ($\theta=\pi/2,\phi=0$) is aligned with the longest principal
    axis. The angular ranges are defined as $0\leq\theta\leq\pi$ and
    $0\leq\phi<2\pi$; however, since projections in opposite
    directions are equivalent in the context of our analysis, we
    consider the coordinates of the set of \nproj\ unique lines of
    sight within the half-sphere defined by $0\leq\theta<\pi$ and
    $0\leq\phi<\pi$ in the grid shown here (extending the grid to
    $\phi=2\pi$ would result in repetition of the data). Since the
    lines of sight result from an evenly distributed spherical
    tessellation, plotting $\phi$ versus $\cos(\theta)=z/r$ ensures
    that the number of projections per grid pixel is approximately
    constant. The $\alpha_\mathrm{Walker}(\theta,\phi)$ value shown in
    each pixel is the mean over all projections of each galaxy within
    the relevant angular range. The colour scale is centred on a mass
    ratio of unity (i.e.\ white for accurate mean estimates, blue for
    mean underestimates, and red for mean overestimates). The results
    for the \citet{Wolf_2010} estimator are very similar to those
    shown here. The distributions of the mass ratios, $\alpha$, for
    both estimators are shown in the upper panel of
    Fig.~\ref{fig:est_Krot} (summing over all projections), and the
    $\alpha$ variations relative to the galactic mean are shown in the
    upper panels of Fig.~\ref{fig:est_angle} (integrating over
    $\theta$ and $\phi$ separately).}
  \label{fig:est_angle_grid}
\end{figure}

\begin{figure}
  \centering
  \includegraphics[width=\columnwidth]{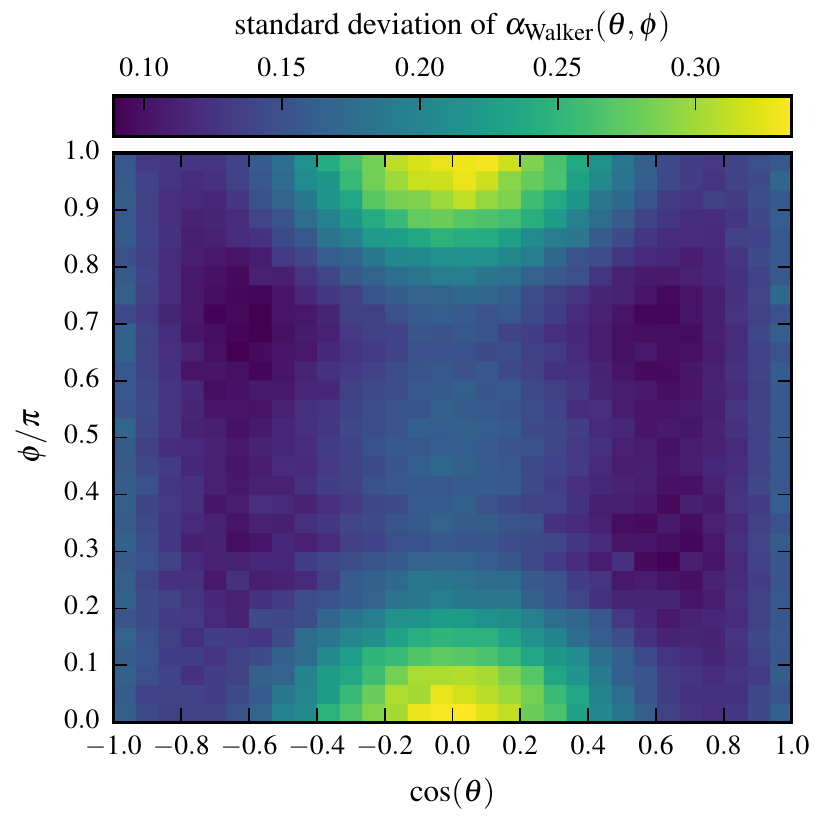}
  \caption{The same as Fig.~\ref{fig:est_angle_grid}, but showing
    instead the standard deviation of the
    $\alpha_\mathrm{Walker}(\theta,\phi)$ values in each pixel, rather
    than the mean value. The details of the analysis are otherwise
    exactly as described in Fig.~\ref{fig:est_angle_grid}.}
  \label{fig:est_angle_grid_sdev}
\end{figure}

\subsection{Stellar velocity dispersion anisotropy}

The anisotropy, $\beta$, of the stellar velocity dispersion is an
important parameter in the Jeans analysis. An important feature of the
mass estimators of interest here is that they are designed to be
minimally sensitive to $\beta$. In Fig.~\ref{fig:est_many}(a), we show
the $\alpha$ distributions for dispersion-dominated galaxies, split
into bins of $\langle\beta\rangle$, which is the stellar velocity
dispersion anisotropy averaged over the whole galaxy (see
equation~\ref{eq:beta}). The stellar orbits are predominantly radially
biased ($\langle\beta\rangle>0$); there are only 46 out of 250
galaxies with $\langle\beta\rangle<0$ (see Fig.~\ref{fig:beta_pro} for
the radial variation of $\beta$ in the HR simulations). For each
estimator, the scatter in $\alpha$ increases with
$\langle\beta\rangle$, where the difference is greatest between the
intermediate and most radially biased systems, for the
$\langle\beta\rangle$ intervals shown.

\subsection{Equilibrium}

In order to assess how the accuracy of the mass estimators depends on
whether the galaxy of interest is in equilibrium, as assumed in the
derivation of each estimator, we make use of $\tilde{\gamma}$ as
defined in Section~\ref{sec:eq_spher}. This encodes the balance
between the total stellar kinetic energy and the potential energy
measured with respect to the bottom of the gravitational potential
well, where galaxies with $\tilde{\gamma}\approx 1$ are assumed to be
in equilibrium (the calculation of $\tilde{\gamma}$ assumes that, on
average, the galaxies in our full sample are in equilibrium; see
Fig.~\ref{fig:mstar_spher_eqrat}). The $\alpha$ distributions for
dispersion-dominated galaxies divided into bins of $\tilde{\gamma}$
are shown in Fig.~\ref{fig:est_many}(b), where galaxies in the
intermediate bin are taken to be close to equilibrium. The scatter in
$\alpha$ actually increases with $\tilde{\gamma}$, such that the
lowest $\tilde{\gamma}$ interval exhibits the smallest scatter for
each estimator.

\subsection{Shape}
\label{sec:est_shape}

As shown in Fig.~\ref{fig:mstar_spher_eqrat}, the simulated galaxies
span a broad range of 3D shapes. Fig.~\ref{fig:est_many}(c) shows the
$\alpha$ distributions for dispersion-dominated galaxies divided into
bins of stellar sphericity, $s$. The scatter in $\alpha$ diminishes
sharply as the stellar mass distribution tends towards spherical
symmetry ($s=1$). This behaviour is consistent with the expectation
for the applicability of the spherical Jeans equation. Clearly, the
estimators perform relatively well for galaxies that are close to
spherical. However, robustly identifying such systems from projected
data alone is non-trivial.

We now investigate the impact of carrying out the shape determination
in projection. Adapting the procedure described in
Section~\ref{sec:eq_spher} for computing sphericity, we compute the
reduced inertia tensor in 2D using the projected coordinates of each
star particle on the sky; the resulting ellipse axis lengths,
$a_\mathrm{p} \geq b_\mathrm{p}$, are then used to define the stellar
circularity as $s_\mathrm{p}=b_\mathrm{p}/a_\mathrm{p}$ (a circle has
$s_\mathrm{p}=1$). Fig.~\ref{fig:est_many}(d) shows the $\alpha$
distributions in bins of circularity, where we see a much weaker trend
in the scatter as a function of shape than when the full 3D
information of Fig.~\ref{fig:est_many}(c) is used.  As such, the
scatter in $\alpha$ for galaxies that appear to be very close to
circular on the sky $(s_\mathrm{p}>0.9)$ is significantly larger than
for galaxies that are actually close to spherical in 3D $(s>0.75)$.

\subsection{Stellar mass}
\label{sec:est_mstar}

The simple mass estimators could be expected to work equally well for
stellar systems on any mass scale, provided that they satisfy the
spherical Jeans equation and that more detailed assumptions relating
to the stellar density distribution and the stellar velocity
dispersion are reasonable (see Section~\ref{sec:est}). However, many
galaxy properties scale with stellar mass, which for our simulated
galaxies, albeit weakly and with a large scatter, include the level of
rotational support and the 3D shape (see Figs.~\ref{fig:Krot_mstar}
and~\ref{fig:mstar_spher_eqrat}). In Fig.~\ref{fig:est_many}(e) we
show the $\alpha$ distributions for dispersion-dominated galaxies in
bins of stellar mass. There is no monotonic global trend in the
scatter in $\alpha$ as a function of stellar mass; however, for both
estimators the scatter peaks in the intermediate mass interval.

\subsection{Structural relationship with dark matter halo}

As discussed above, Figs.~\ref{fig:est_Krot} and~\ref{fig:est_many}
explore the dependence of the mass estimator errors on key structural
and kinematical properties of the galaxy itself, i.e.\ properties of
the set of subhalo star particles enclosed within a sphere of radius
$r_\mathrm{gal}$ around the subhalo centre (including the balance
between the stellar kinetic energy and the gravitational potential
energy of the stars due to the mass distribution of the whole
subhalo).  Extending this, we now examine the dependence of the mass
estimator accuracy on quantities that characterise the structural
relationship between the galaxy and the dark matter component of its
host subhalo.

Fig.~\ref{fig:est_many_halo}(a) shows the $\alpha$ distributions for
dispersion-dominated galaxies, divided into bins according to the
angle, $\psi$, between the minor axis of the galaxy and that of the
dark matter component of its subhalo. For each mass component, the
minor axis is taken to be given by the eigenvector of the reduced
inertia tensor that has the smallest eigenvalue, and thus corresponds
to the shortest axis length (i.e.\ $c$ in $s=c/a$; see
Section~\ref{sec:eq_spher}). Note that the signs of the eigenvectors
are ignored, such that the angle $\psi$ can take values from zero to
$\pi/2$. The scatter in $\alpha$ is highest for the galaxies that are
close to perfect alignment with their dark matter halo ($\psi=0$), and
the scatter decreases with increasing $\psi$.

In order to investigate the dependence of the estimator accuracy on
how deeply the galaxy is embedded within its host subhalo, we compute
a scale radius, $r_\mathrm{DM}$, for the subhalo dark matter mass
distribution and compare this to the stellar 3D half-mass radius,
$r_\mathrm{half}$. Given the radius at which the circular velocity
curve peaks, considering the enclosed mass in subhalo dark matter
particles \textit{only}, $r_\mathrm{DM}$ is computed as the
corresponding scale radius for an exact NFW profile
\citep{NFW_1996}.\footnote{Given the enclosed dark matter mass
  profile, $M_\mathrm{DM}(<r)$, if the radius at which the
  \textit{dark matter only} circular velocity curve,
  $\sqrt{GM_\mathrm{DM}(<r)/r}$, peaks is $r_\mathrm{max(DM)}$, then
  the scale radius for an exact NFW halo with the same value of
  $r_\mathrm{max(DM)}$ is
  $r_\mathrm{DM} \simeq r_\mathrm{max(DM)} / 2.163$.}
Fig.~\ref{fig:est_many_halo}(b) shows the $\alpha$ distributions for
dispersion-dominated galaxies in bins of
$r_\mathrm{half}/r_\mathrm{DM}$, which quantifies the `concentration'
of the galaxy with respect to the dark matter distribution. The
scatter in $\alpha$ peaks in the intermediate bin shown, reducing
somewhat in the most concentrated bin (lowest
$r_\mathrm{half}/r_\mathrm{DM}$), followed by the least concentrated
systems, but the differences are small.

Finally, it is interesting to examine the dependence of the estimator
error on the 3D shape of the host dark matter halo, since the mass
estimators assume that the dynamical system is spherically symmetric
as a whole, and the scatter in $\alpha$ is found to be strongly
dependent on the 3D shape of the stellar distribution (see
Fig.~\ref{fig:est_many}c). We compute the sphericity of the subhalo
dark matter distribution, $s_\mathrm{DM}$, in the same way as
described in Section~\ref{sec:eq_spher} for the stellar sphericity,
$s$. Fig.~\ref{fig:est_many_halo}(c) shows the $\alpha$ distributions
for dispersion-dominated galaxies in bins of $s_\mathrm{DM}$. The
scatter in $\alpha$ decreases with increasing $s_\mathrm{DM}$; however
the strength of this dependence is smaller than that for the stellar
sphericity, comparing to Fig.~\ref{fig:est_many}(c).  This indicates
that the shape of the stellar mass distribution has the largest impact
on the estimator error, with a weaker influence from the shape of the
mass distribution of the dark matter.  We note that although $s$ does
exhibit a weak increase on average as a function of $s_\mathrm{DM}$,
there is a very large scatter in $s$ at a given $s_\mathrm{DM}$, i.e.\
the stellar and dark matter 3D shapes are only loosely connected. The
dark matter distribution is usually closer than the galaxy to being
spherically symmetric (only 24 out of 250 galaxies have
$s > s_\mathrm{DM}$).  Fig.~\ref{fig:est_many_halo}(d) shows the
$\alpha$ distributions for dispersion-dominated galaxies in bins of
$s/s_\mathrm{DM}$. The decrease in the scatter in $\alpha$ is more
significant as a function of this ratio than as a function of
$s_\mathrm{DM}$ (cf.\ Fig.~\ref{fig:est_many_halo}c), but less
pronounced than as a function of $s$ alone (cf.\
Fig.~\ref{fig:est_many}c).  Therefore, the difference between the 3D
shapes of the galaxy and its host dark matter halo (quantified here by
$s/s_\mathrm{DM}$) plays a role in the accuracy of simple mass
estimators.

\subsection{Field versus satellite galaxies}

Fig.~\ref{fig:est_type} shows the $\alpha$ distributions separately
for dispersion-dominated field and satellite galaxies (satellites are
within $300\,\mathrm{kpc}$ of a MW or M31 analogue). Although there
are fewer satellites, and hence poorer statistics, the scatter in
$\alpha$ is smaller for the satellite galaxies than for the field
galaxies, for both estimators. This may be closely related to the fact
that, on average, the satellite galaxies are closer to spherical
symmetry than those in the field (see
Fig.~\ref{fig:mstar_spher_eqrat}). There are hints that the satellites
also tend to be more strongly supported by dispersion (see
Fig.~\ref{fig:Krot_mstar}) and tend to have stellar velocity
dispersions that are closer to isotropic (see Fig.~\ref{fig:beta_pro}
for HR), compared with field galaxies. We also find that the satellite
galaxies tend to be more extended relative to the dark matter
distribution of their subhalo (higher $r_\mathrm{half}/r_\mathrm{DM}$)
than the field galaxies.

\subsection{Angular dependence}
\label{sec:est_angle}

The analysis presented so far has considered the estimator error
distributions resulting from summing over all galaxies (or lines of
sight) satisfying certain criteria based on integrated stellar and
dark matter properties (or per-projection observable properties). We
now investigate how the estimator accuracy varies with viewing angle
relative to the orientation of the galaxy. Given Cartesian coordinates
($x,y,z$), we adopt the standard convention for the spherical polar
coordinates ($r,\theta,\phi$), such that
$x = r \sin(\theta) \cos(\phi)$, $y = r \sin(\theta) \sin(\phi)$, and
$z = r \cos(\theta)$. For each galaxy, we rotate the simulation
coordinate system to align with the eigenvectors of the stellar
reduced inertia tensor (see Section~\ref{sec:eq_spher}). The $z$ axis
($\theta=0$) is aligned with the shortest principal axis (minor axis),
the $x$ axis ($\theta=\pi/2,\phi=0$) with the longest principal axis
(major axis), and the remaining principal axis with the $y$ direction
($\theta=\pi/2,\phi=\pi/2$).

The upper panels of Fig.~\ref{fig:est_angle} show the variation in the
$\alpha$ values for each estimator, relative to the mean $\alpha$
value from averaging over all lines of sight for a given galaxy, as a
function of $\theta$ and $\phi$, defined using the \textit{stellar}
reduced inertia tensor.  The angular variations of the projected
half-mass radius, $R_\mathrm{e}$, and the line-of-sight integrated
stellar velocity dispersion, $\langle\sigma_\mathrm{los}\rangle$, are
also shown, again relative to the corresponding galactic means.  The
lower panels in Fig.~\ref{fig:est_angle} repeat the same analysis, but
using instead the $\theta$ and $\phi$ coordinates with respect to the
eigenbasis of the reduced inertia tensor of the \textit{dark matter}
particles within the subhalo. Thus, for a given galaxy, the
($\theta,\phi$) coordinates contributing to the upper and lower panels
of Fig.~\ref{fig:est_angle} differ unless the stellar and dark matter
reduced inertia tensor eigenvectors are exactly aligned, with the same
eigenvalue ordering (since this sets the ordering of the principal
axis lengths, and hence the definitions of $\theta$ and $\phi$).

Looking at the upper panels of Fig.~\ref{fig:est_angle}, it is clear
that $\alpha$ is strongly sensitive to $\theta$, i.e.\ the angle with
respect to the stellar minor axis ($z$), and has a weaker dependence
on $\phi$, i.e.\ the angle with respect to the major axis ($x$; in the
$x-y$ plane). The results for the two different estimators are very
similar, although the estimator of \cite{Wolf_2010} exhibits a
slightly smaller scatter at all angles.  As a function of $\theta$
(upper left panel), the mass estimates are maximally biased low (in
the median for all galaxies), relative to the galactic mean values,
for lines of sight coincident with the stellar minor axis
($\theta\sim0$ and $\theta\sim\pi$).  The median estimates then
smoothly increase towards the $x-y$ plane, peaking for
$\theta/\pi\sim0.5$. The scatter in $\alpha$, relative to the galactic
mean, is relatively large for lines of sight along the minor axis or
within the $x-y$ plane, but is minimised for $\theta/\pi\sim0.3$ and
$\theta/\pi\sim0.7$, being approximately symmetrical around
$\theta/\pi=0.5$. The scatter in both $R_\mathrm{e}$ and
$\langle\sigma_\mathrm{los}\rangle$ exhibits the same behaviour as a
function of $\theta$ as the scatter in $\alpha$, with extremes at
approximately the same values of $\theta$.  The velocity dispersion
varies with $\theta$ in the same way as $\alpha$ does, peaking for
$\theta/\pi\sim0.5$, with minima coinciding with the minor
axis. However, the dependence of $R_\mathrm{e}$ on $\theta$ is
reversed, compared to that of $\alpha$ and
$\langle\sigma_\mathrm{los}\rangle$, such that the $R_\mathrm{e}$
curve is approximately the reflection of the
$\langle\sigma_\mathrm{los}\rangle$ curve around the galactic mean.
The coupled variation of $R_\mathrm{e}$ and
$\langle\sigma_\mathrm{los}\rangle$ suppresses the variation in the
product $R_\mathrm{e}\langle\sigma_\mathrm{los}\rangle^2$, and hence
the variation in the estimated mass (see
equations~\ref{eq:alpha_walker} and~\ref{eq:alpha_wolf}).  The
$\alpha$, $R_\mathrm{e}$, and $\langle\sigma_\mathrm{los}\rangle$
curves intersect (with values $\sim 1$) for $\theta$ close to where
the scatter in each quantity is minimised ($\theta/\pi\sim0.3$ and
$\theta/\pi\sim0.7$).

The results as a function of $\phi$ in galactic coordinates (upper
right panel in Fig.~\ref{fig:est_angle}) are similar to those just
described for the $\theta$ dependence, but with smaller median offsets
from unity and larger scatters, exhibiting extremes such that $\alpha$
and $\langle\sigma_\mathrm{los}\rangle$ are maximised for $\phi\sim0$
and $\phi\sim\pi$ (i.e.\ for lines of sight in the $x-z$ plane), and
minimised for $\phi/\pi\sim0.5$ (i.e.\ the $y-z$ plane); while the
$\phi$ values of the turning points are shifted by an angle of $\pi/2$
relative to this for $R_\mathrm{e}$. Note that since projections in
opposite directions are equivalent for our purposes, the angular
coordinates of the \nproj\ unique lines of sight are evaluated within
the half-sphere defined by $0\leq\theta<\pi$ and $0\leq\phi<\pi$ in
Fig.~\ref{fig:est_angle}.

Using instead the coordinate system based on the subhalo dark matter,
as shown in the lower panels of Fig.~\ref{fig:est_angle}, results in
dependencies of $\alpha$, $R_\mathrm{e}$, and
$\langle\sigma_\mathrm{los}\rangle$ on the spherical polar angles that
are qualitatively very similar to the data shown in the upper panels,
where the coordinates based on the galactic star particles are
used. However, the strengths of the trends seen in galactic
coordinates are diluted when switching to the halo-based coordinates,
such that the median variations are reduced for each quantity shown
(i.e.\ the solid lines are slightly closer to unity).  The scatter
around the median also increases for each quantity at all angles when
switching to the halo coordinates, except for $\phi/\pi\lesssim0.2$
and $\phi/\pi\gtrsim0.8$, where for each curve shown, the
$16^\text{th}-84^\text{th}$ percentile width actually decreases. The
differences between the upper and lower panels in
Fig.~\ref{fig:est_angle} indicate that the accuracy of the mass
estimators, for a given line of sight, is more sensitive to the shape
and relative alignment of the galaxy, rather than the corresponding
properties of the dark matter (sub)halo in which the galaxy is
embedded.

Fig.~\ref{fig:est_angle_grid} shows the mean $\alpha$ value for the
\cite{Walker_2009} estimator, as a function of $\cos(\theta)$ and
$\phi$, for all dispersion-dominated galaxies, using the coordinate
system based on the reduced inertia tensor of the stars within each
galaxy, as in the upper panels of Fig.~\ref{fig:est_angle} (but note
that Fig.~\ref{fig:est_angle} considers the $\alpha$ values relative
to the galactic mean, for each galaxy, while in
Fig.~\ref{fig:est_angle_grid} the actual individual $\alpha$ values
are used directly to compute the mean for each pixel). Since the lines
of sight are drawn from an evenly spaced spherical tessellation, they
must evenly sample the $\cos(\theta)$ versus $\phi$ plane, and so the
pixels in Fig.~\ref{fig:est_angle_grid} each consider approximately
the same number of data points (galaxy projections). The results for
the \cite{Wolf_2010} estimator are very similar to those shown in
Fig.~\ref{fig:est_angle_grid}.  As can be seen in
Fig.~\ref{fig:est_angle_grid}, the most accurate mean mass estimates
form an hourglass shape in the $\cos(\theta)-\phi$ plane (white). The
masses are most heavily underestimated for $\cos(\theta)\sim\pm1$
(blue; corresponding to lines of sight along the stellar
\textit{minor} axis), and most heavily overestimated for
$\cos(\theta)\sim0$ with $\phi/\pi\sim0$ or $\phi/\pi\sim1$ (red;
corresponding to lines of sight along the stellar \textit{major}
axis). Lines of sight coincident with the stellar principal axis of
\textit{intermediate} length, where $\cos(\theta)=0$ and
$\phi/\pi=0.5$ (or $\phi/\pi=1.5$; not shown), correspond to the
centre of the grid shown in Fig.~\ref{fig:est_angle_grid}, where the
mean mass estimates are relatively close to perfect accuracy.

Fig.~\ref{fig:est_angle_grid_sdev} is the same as
Fig.~\ref{fig:est_angle_grid}, except that it shows the standard
deviation of the $\alpha$ values, as a function of $\cos(\theta)$ and
$\phi$; otherwise the calculations involved are identical for the two
figures. It can be seen in Fig.~\ref{fig:est_angle_grid_sdev} that the
scatter in $\alpha$ is highest for lines of sight along the stellar
major axis, and lowest for the two regions with
$\cos(\theta)\sim\pm0.6$ (i.e.\ $\theta/\pi\sim0.3$ and
$\theta/\pi\sim0.7$) and $0.2\lesssim\phi/\pi\lesssim0.8$ (in keeping
with the upper panels of Fig.~\ref{fig:est_angle}).  The scatter for
lines of sight coincident with the stellar minor axis and the
intermediate stellar principal axis is fairly similar, in stark
contrast to the major axis. Comparing Figs.~\ref{fig:est_angle_grid}
and~\ref{fig:est_angle_grid_sdev}, we see that the locations of the
largest scatter in $\alpha$ are also the locations of the highest mean
$\alpha$ values. For the three stellar principal axes, the dynamical
masses are either overestimated with a large scatter (major axis),
underestimated with a modest scatter (minor axis), or in fact close to
accurate with a modest scatter (intermediate axis).

\begin{figure*}
  \centering
  \includegraphics[width=\textwidth]{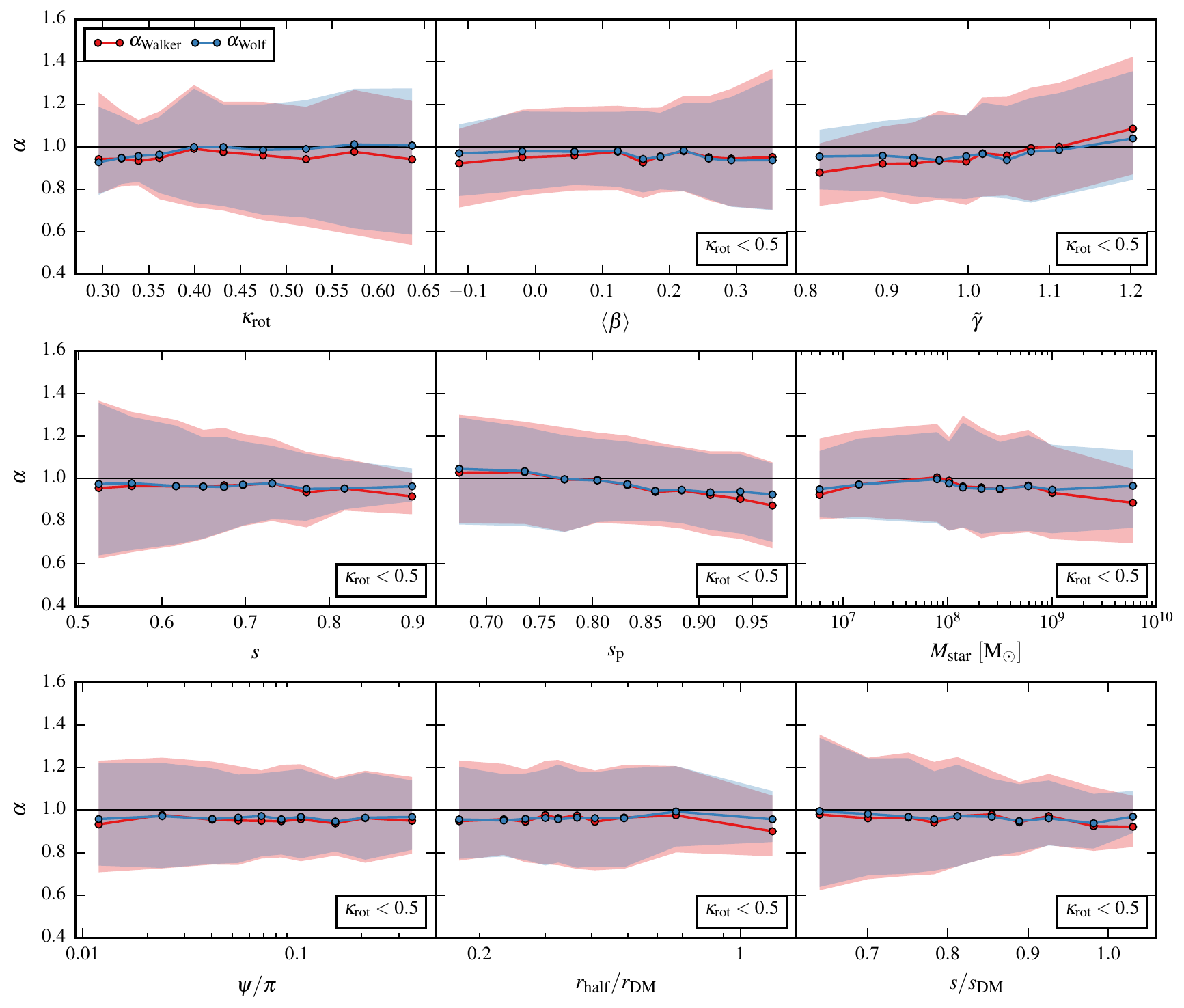}
  \caption{Ratio, $\alpha$, of the estimated to true mass obtained
    using the estimators of
    \citet[$\alpha_\mathrm{Walker}$]{Walker_2009} and
    \citet[$\alpha_\mathrm{Wolf}$]{Wolf_2010}, by projecting over
    \nproj\ evenly distributed lines of sight, as a function of
    various galaxy and subhalo dark matter properties. The properties
    shown are: the stellar kinematic measure used to discriminate
    between galaxies with different levels of rotational support
    ($\kappa_\mathrm{rot}$), the mean stellar velocity dispersion
    anisotropy ($\langle\beta\rangle$), the stellar equilibrium
    measure ($\tilde{\gamma}$), the stellar sphericity ($s$), the
    projected stellar circularity ($s_\mathrm{p}$), the stellar mass
    ($M_\mathrm{star}$), the angle between the stellar and subhalo
    dark matter minor axes ($\psi$), the ratio of the stellar and dark
    matter scale radii ($r_\mathrm{half}/r_\mathrm{DM}$), and the
    ratio of the stellar and dark matter sphericities
    ($s/s_\mathrm{DM}$). The first panel shows $\alpha$ versus
    $\kappa_\mathrm{rot}$ for all galaxies; all other panels consider
    only those galaxies whose stellar motions are dominated by
    dispersion ($\kappa_\mathrm{rot}<0.5$). For each estimator
    (colours), the solid lines show the median mass ratios and the
    shaded regions the $16^\text{th}-84^\mathrm{th}$ percentile
    spread. The bin edges are evenly spaced percentiles of the
    quantity on the horizontal axis, and the bin centres (points) are
    the median values in each bin.}
  \label{fig:est_summary}
\end{figure*}

\subsection{Summary and discussion}
\label{sec:est_summary}

To summarise the preceding results, Fig.~\ref{fig:est_summary} shows
$\alpha$ for each estimator as a function of various galaxy
properties, in bins of approximately equal numbers of galaxies (or
sight-lines). The first panel shows the dependence on
$\kappa_\mathrm{rot}$ for the full galaxy sample, and the subsequent
panels consider only the dispersion-dominated galaxies
($\kappa_\mathrm{rot}<0.5$). The distinction between the two
estimators is small, although where they deviate, the
\citet{Wolf_2010} estimator has the smallest bias and scatter in
almost all cases.

Considering the scatter, the estimators are most precise for
dispersion-supported galaxies that are close to spherical, have at
most mildly radially biased stellar velocity dispersions
($\langle\beta\rangle\lesssim 0.2$), and have relatively low stellar
kinetic energy for their gravitational potential energy. The strongest
dependence of the scatter is on the 3D shape of the stellar mass
distribution, of which almost all useful information is lost in
projection. Considering only the dispersion-dominated galaxies, as
shown in Fig.~\ref{fig:est_summary}, the scatter also reduces for
galaxies that are misaligned with respect to their host dark matter
halo, and for galaxies that are highly extended, in units of their
halo scale radius. It is interesting to note that the
(dispersion-dominated) satellite galaxies exhibit a lower $\alpha$
scatter overall compared to their counterparts in the field (see
Fig.~\ref{fig:est_type}). Satellite galaxies also tend to be closer to
spherical and tend to be more extended relative to their host dark
matter halo, compared to the field galaxies (the highest
$r_\mathrm{half}/r_\mathrm{DM}$ bin in Fig.~\ref{fig:est_summary}
contains 21 satellite galaxies and only four field galaxies). We note
that the main physical galaxy properties of interest are related at
some level: galaxies that are close to spherical tend to have velocity
dispersions that are close to isotropic and have relatively little
rotational support. The scatter in $\alpha$ is much more sensitive to
the shape of the galaxy than to that of the mass distribution of the
dark matter within its host subhalo, such that there remain strong
trends in the scatter in $\alpha$ as a function of the ratio of
stellar and dark matter sphericities, $s/s_\mathrm{DM}$.

Any trends in the median $\alpha$ values shown in
Fig.~\ref{fig:est_summary} are weak. However, considering the
dependence on the projected stellar circularity, $s_\mathrm{p}$, the
bias towards overestimation of the mass for galaxies that are very
elongated on the sky gradually decreases with increasing circularity
(we note that this trend is exaggerated if we instead consider the
full galaxy sample, including the rotation-dominated galaxies). The
median $\alpha$ also exhibits an increase with increasing
$\tilde{\gamma}$ (i.e.\ for increasing stellar kinetic energy relative
to the stellar gravitational potential energy, at fixed stellar mass).

As shown in Fig.~\ref{fig:est_angle} (upper panels), the value of
$\alpha$ is strongly sensitive to the viewing angle, $\theta$,
relative to the stellar minor axis, $z$, such that the estimated
masses are lowest for lines of sight close to the minor axis, and
highest for lines of sight in the $x-y$ plane. The dependence of
$\alpha$ on the other spherical polar viewing angle, $\phi$, relative
to the stellar major axis, $x$ (in the $x-y$ plane), is much weaker
than the dependence on $\theta$, but shows similar behaviour. As shown
in Figs.~\ref{fig:est_angle_grid} and~\ref{fig:est_angle_grid_sdev},
the dynamical masses are overestimated with a large scatter for lines
of sight coincident with the major axis, underestimated with a modest
scatter for the minor axis, and close to accurate with a modest
scatter for the intermediate principal axis.

We note that \citet{Wolf_2010} suggest an error of around
$0.05\,\mathrm{dex}$ (12 percent; cf.\ Fig.~\ref{fig:est_summary}) for
their estimator (in addition to the uncertainty due to the errors on
the measured size and velocity dispersion), in the special case where
the unknown $\beta(r)$ has an extremum within the stellar
distribution, or $\sigma_\mathrm{los}(R)$ has not been measured over
the full extent of the galaxy (i.e.\ the dispersion much beyond
$R \sim R_\mathrm{e}$ is not included when
$\langle\sigma_\mathrm{los}\rangle$ is computed). Otherwise,
\citet{Wolf_2010} advocate considering only the measurement errors on
the velocity dispersion and half-light radius when evaluating the
uncertainties in masses obtained using their estimator.

\citet{Kow_2013} apply a modified version of the estimator of
\citet{Wolf_2010} to idealised simulations of dSphs in a static MW
potential, and find that the error decreases as the shape of the
stellar mass distribution tends towards spherical symmetry, as also
found here. However, they also find that for the simulated dSphs there
is no clear dependence of the error on the level of stellar rotation
or velocity dispersion anisotropy, while we do detect such
dependencies here (see Fig.~\ref{fig:est_summary}).

\citet{Laporte_2013_A} investigate the accuracy of the
\citet{Walker_2009} estimator for idealised stellar distributions
placed in dark matter subhaloes from the Aquarius simulations, finding
fluctuations of between 10 and 20 percent with respect to the true
mass. An interesting result of this study, which naturally takes into
account the triaxiality of dark matter haloes formed in cosmological
simulations, is that it finds that the enclosed dynamical mass is more
strongly overestimated for stellar populations that are more deeply
embedded (more highly concentrated) within their dark matter halo,
concurring with the spherically symmetric tests of
\cite{Walker_2011}. However, we do not detect significant evidence of
such a trend in the APOSTLE simulations (see the
$r_\mathrm{half}/r_\mathrm{DM}$ panel of Fig.~\ref{fig:est_summary}).
Additionally, \citet{Laporte_2013_A} demonstrate variations of the
estimated mass, $R_\mathrm{e}$, and
$\langle\sigma_\mathrm{los}\rangle$, as a function of viewing angle,
that are in good qualitative agreement with the results presented in
Section~\ref{sec:est_angle}.

\citet{Lyskova_2015} apply the \citet{Wolf_2010} estimator to
cosmological zoom simulations of 40 isolated elliptical galaxies,
finding a $1\sigma$ scatter of around 8 percent for
$V_\mathrm{c}(4R_\mathrm{e}/3)$ (note that the fractional error on
$V_\mathrm{c}(r)$ is half that on $M(<r)$, assuming zero error on the
radius), with a bias of 3 percent above the true circular velocity
(cf.\ Fig.~\ref{fig:est_summary}).

Our results complement these earlier studies, but we have used a large
sample of realistic galaxies with a broad range of stellar masses,
formed in a self-consistent $\Lambda$CDM cosmological context, in the
environment of the Local Group, using sophisticated treatments of the
baryonic physics important for galaxy formation.

\section{Effects of uncertainties}
\label{sec:effects}

We now explore the impact of the systematic uncertainties associated
with the simple mass estimators on key results in the literature that
are based on applying the estimators to observational data.  Despite
the high resolution of the APOSTLE simulations, our sample of galaxies
does not include systems with stellar masses smaller than that of
Sculptor (due to our chosen resolution threshold; see
Section~\ref{sec:conv} and Fig.~\ref{fig:rhalf_mstar_sig}). However,
the estimator error has no clear dependence on stellar mass (see
Fig.~\ref{fig:est_many}e). Therefore, we assume that the uncertainties
explored in Section~\ref{sec:app} remain relevant for galaxies that
are fainter than those in our sample.

\subsection{Dynamical masses of MW dSphs}

\citet{Wolf_2010} find that simply propagating the observational
errors on the half-light radius and the line-of-sight velocity
dispersion through their estimator equation yields a similar
uncertainty for the estimated dynamical mass to that obtained from
their full Jeans analysis (which, like the estimator, assumes
spherical symmetry). The authors argue that this is consistent with
their claim that the uncertainty on the recovered mass is dominated by
observational errors, rather than underlying systematic
effects. However, as we have demonstrated, the estimator's precision
is sensitive to various properties of the target galaxy.

To illustrate the significance of the systematic error on the
estimated mass, Table~\ref{tab:wolf_data} lists the mass within the 3D
half-light radius for dSphs of the MW, which we compute using
equation~(\ref{eq:wolf2D}), given the projected half-light radius and
line-of-sight velocity dispersion values tabulated by
\citet{Wolf_2010}.\footnote{Note that \citet{Wolf_2010} tabulate the
  mass within the 3D half-light radius from their full Jeans analysis,
  rather than from their estimator.}  For each galaxy, we list the
$1\sigma$ fractional error, $\sigma_M$, on the estimated mass,
obtained by propagating the observational errors through
equation~(\ref{eq:wolf2D}), as advocated by \citet{Wolf_2010}.  The
final column gives the fractional error on the mass if, in addition,
we add in quadrature a \textit{representative} $1\sigma$ systematic
error of 20 percent in the estimated to true mass ratio.  This
fiducial value for the scatter is appropriate on average for the
dispersion-dominated satellite galaxies in the simulations (see
Fig.~\ref{fig:est_type}), or all dispersion-dominated galaxies that
appear near-circular on the sky (see Fig.~\ref{fig:est_many}d). For
many of the bright satellites, the observational errors on the
measured quantities are sufficiently small that the increase in the
overall mass error due to including the representative systematic
uncertainty is similar in size to the original mass error. However, in
many other cases the original error is so large that the contribution
from the systematic uncertainty is relatively insignificant.

\begin{table}
  \centering
  \caption{Dynamical mass, $M_\mathrm{Wolf}$, within the 3D half-light
    radius ($r_\mathrm{half}\approx 4R_\mathrm{e}/3$) of dSph
    satellites of the MW, according to the estimator of
    \citet{Wolf_2010}, as given in equation~(\ref{eq:wolf2D}). The
    estimates have been computed using the projected half-light radii,
    $R_\mathrm{e}$ (as reproduced here), and line-of-sight velocity
    dispersions tabulated by \citet{Wolf_2010}.  $\sigma_M$ is the
    $1\sigma$ fractional error on $M_\mathrm{Wolf}$, obtained by
    propagating the observational uncertainties on the size and
    dispersion measurements through equation~(\ref{eq:wolf2D}),
    approximating the errors as symmetric.  To illustrate the impact
    of the systematic errors explored in this paper, the final column
    gives the fractional error,
    $\tilde{\sigma}_M=(\sigma_M^2+\sigma_\mathrm{sys}^2)^{1/2}$, on
    the estimated mass if a \textit{representative} systematic error
    of $\sigma_\mathrm{sys}=0.2$ is assumed for the estimated to true
    mass ratio (ignoring any bias; see Fig.~\ref{fig:est_type}).  The
    galaxies are listed in order of decreasing $M_\mathrm{Wolf}$.}
  \label{tab:wolf_data}
  \begin{tabular}{lcccc}
    \hline
    Galaxy
    & $R_\mathrm{e}~[\mathrm{pc}]$
    & $\log_{10}(M_\mathrm{Wolf}~[\mathrm{M}_\odot])$
    & $\sigma_M$
    & $\tilde{\sigma}_M$ \\
    \hline
    Fornax             & $714_{-40}^{+40}$ & 7.88 & 0.07 & 0.21 \\
    Ursa Minor         & $445_{-44}^{+44}$ & 7.74 & 0.14 & 0.25 \rule{0pt}{2.5ex} \\
    Sextans            & $768_{-47}^{+47}$ & 7.56 & 0.10 & 0.23 \rule{0pt}{2.5ex} \\
    Canes Venatici I   & $564_{-36}^{+36}$ & 7.48 & 0.15 & 0.25 \rule{0pt}{2.5ex} \\
    Leo I              & $295_{-49}^{+49}$ & 7.35 & 0.19 & 0.27 \rule{0pt}{2.5ex} \\
    Sculptor           & $282_{-41}^{+41}$ & 7.33 & 0.15 & 0.25 \rule{0pt}{2.5ex} \\
    Draco              & $220_{-11}^{+11}$ & 7.32 & 0.11 & 0.23 \rule{0pt}{2.5ex} \\
    Bootes I           & $242_{-20}^{+22}$ & 7.26 & 0.50 & 0.54 \rule{0pt}{2.5ex} \\
    Ursa Major I       & $318_{-39}^{+50}$ & 7.23 & 0.30 & 0.36 \rule{0pt}{2.5ex} \\
    Carina             & $254_{-28}^{+28}$ & 6.99 & 0.13 & 0.24 \rule{0pt}{2.5ex} \\
    Leo II             & $177_{-13}^{+13}$ & 6.86 & 0.17 & 0.26 \rule{0pt}{2.5ex} \\
    Leo T              & $115_{-17}^{+17}$ & 6.81 & 0.44 & 0.48 \rule{0pt}{2.5ex} \\
    Ursa Major II      & $140_{-25}^{+25}$ & 6.77 & 0.45 & 0.50 \rule{0pt}{2.5ex} \\
    Hercules           & $229_{-19}^{+19}$ & 6.74 & 0.36 & 0.41 \rule{0pt}{2.5ex} \\
    Coma Berenices     & $77_{-10}^{+10}$ & 6.18 & 0.37 & 0.42 \rule{0pt}{2.5ex} \\
    Canes Venatici II  & $74_{-10}^{+14}$ & 6.16 & 0.46 & 0.51 \rule{0pt}{2.5ex} \\
    Leo IV             & $116_{-34}^{+26}$ & 6.07 & 1.06 & 1.08 \rule{0pt}{2.5ex} \\
    Segue 1            & $29_{-5}^{+8}$ & 5.70 & 0.56 & 0.59 \rule{0pt}{2.5ex} \\
    Willman 1          & $25_{-6}^{+5}$ & 5.57 & 0.50 & 0.54 \rule{0pt}{2.5ex} \\
    \hline
  \end{tabular}
\end{table}

\subsection{Density profiles of Sculptor and Fornax}

\citet{Walker_2011} employ a Markov Chain Monte Carlo (MCMC) technique
to infer projected half-light radii and velocity dispersions for
chemo-dynamically distinct stellar subpopulations in the Sculptor and
Fornax dSphs. In each galaxy there is a centrally concentrated,
kinematically cold, metal-rich stellar component (population 1), and a
more extended, metal-poor component, with higher velocity dispersion
(population 2). The procedure used by \citet{Walker_2011} treats each
galaxy as a superposition of two spherically symmetric stellar
populations (each described by a Plummer profile), with Gaussian
velocity and metallicity distributions. Assuming that the two
populations independently trace the gravitational potential, the
resulting size and dispersion values define a mass slope,
$\Gamma = \Delta \log M(<r) / \Delta \log r$, using
equation~(\ref{eq:est}), where the actual values of $\lambda$ and
$\mu$ are not relevant for computing the slope:
\begin{equation}
  \Gamma = \frac{\log[ M(<r_2) / M(<r_1) ]}{\log [r_2 / r_1]}
  \approx 1 + \frac{\log [ \sigma_2^2 / \sigma_1^2 ]}{\log [r_2 / r_1]}~.
  \label{eq:slope}
\end{equation}
Here $r_1$ and $r_2$ are 3D radii equal to (or some multiple of) the
projected half-light radii, $R_\mathrm{e}$, of populations 1 and 2,
and $\sigma_1$ and $\sigma_2$ are the mean velocity dispersions,
$\langle\sigma_\mathrm{los}\rangle$, of each population, respectively.
Building a posterior probability distribution function, $P(\Gamma)$,
using the set of $r_1$, $r_2$, $\sigma_1$, and $\sigma_2$ values at
each point in the MCMC chains, \citet{Walker_2011} find median slopes
of $\Gamma=2.61^{+0.43}_{-0.37}$ for Fornax and
$\Gamma=2.95^{+0.51}_{-0.39}$ for Sculptor, where the ranges indicate
the $16^\text{th}-84^\text{th}$ percentile confidence intervals.

A measurement of $\Gamma$ using the masses enclosed at two non-zero
radii places an \textit{upper limit} on the inner logarithmic slope,
$\gamma_\mathrm{DM}$, of the dark matter density profile, such that
$\gamma_\mathrm{DM}<3-\Gamma$. Given that $\gamma_\mathrm{DM}=1$ for
an NFW density profile \citep{NFW_1996}, while e.g.\
$\gamma_\mathrm{DM}=0$ for a constant density core,
\citet{Walker_2011} exclude NFW (or steeper) profiles with confidence
of (at least) 95.9 and 99.8 percent for Fornax and Sculptor
respectively. These significance values, $s(\gamma_\mathrm{DM})$, are
computed as,
\begin{equation}
  s(\gamma_\mathrm{DM}) = \frac{\int_{3-\gamma_\mathrm{DM}}^\infty P(\Gamma) \mathrm{d} \Gamma}{\int_{-\infty}^\infty P(\Gamma) \mathrm{d} \Gamma} ~.
  \label{eq:signif}
\end{equation}

If the estimates of the enclosed dynamical masses for the two stellar
populations each happen to be biased by exactly the same (arbitrary)
factor, with respect to the corresponding true masses, then clearly
such a bias cancels out in the calculation of $\Gamma$, i.e.\ the
recovered mass profile slope is insensitive to any coherent bias in
the estimates of the enclosed masses at the two radii (see
equation~\ref{eq:slope}). Therefore, only \textit{differences} in the
enclosed mass estimation biases for the two populations are relevant
in the context of the methodology of \cite{Walker_2011}.

\citet{Laporte_2013_A} generate stellar distribution functions in dark
matter subhaloes drawn from the Aquarius simulations, in order to
assess the sensitivity of the accuracy of the dual population method
of \citet{Walker_2011} to the triaxiality of the gravitational
potential (i.e.\ the lack of spherical symmetry in the dark matter
distribution). The lack of symmetry tends to introduce an
anticorrelation between $R_\mathrm{e}$ and
$\langle\sigma_\mathrm{los}\rangle$ that suppresses the error on the
recovered mass (see also Fig.~\ref{fig:est_angle} and
Section~\ref{sec:wolf3D_true}). \citet{Laporte_2013_A} argue that the
level of spherical symmetry is not important to the estimator
accuracy, in the content of the analysis of \citet{Walker_2011},
despite the fact that they obtain fluctuations of between 10 and 20
percent for the estimated masses (cf.\ Fig.~\ref{fig:est_summary}).
These authors find, as in the spherically symmetric tests of
\citet{Walker_2011}, that $\Gamma$ tends to be systematically
\textit{underestimated}, and thus the exclusion confidences for NFW
profiles in Sculptor and Fornax are deemed to be \textit{conservative}
(i.e.\ the results for $\Gamma$ can be considered as reliable lower
limits, corresponding to reliable upper limits on the density slope,
since $\gamma_\mathrm{DM}<3-\Gamma$).  This is because the enclosed
mass tends to be more strongly overestimated for tracer populations
that are more deeply embedded.  However, as noted in
Section~\ref{sec:est_summary}, we do not find any significant evidence
of such a trend in the bias in the estimated to true mass ratio
($\alpha$) as a function of the concentration of the stellar
distribution ($r_\mathrm{half}/r_\mathrm{DM}$) for the
dispersion-dominated galaxies within our simulations (see
Fig.~\ref{fig:est_summary}).

An important aspect of the tests of \citet{Laporte_2013_A}, which use
idealised independent stellar distributions placed in realistic dark
matter haloes, is that the shapes (equidensity surfaces) of the
stellar populations exactly trace the gravitational potential, and so
the shapes of any two stellar populations generated in this way are
inherently closely correlated, by construction
\citep{Laporte_2013_B}. This close coupling assumed between the two
populations may produce artificially small errors on the recovered
mass slope, as it does not take into account the possibility that the
stellar mass may be distributed in a way that does not exactly follow
the contours in the potential, and that the two populations may have
quite different 3D shapes and relative alignments.

\citet{Kow_2013} carry out a similar study to that of
\citet{Laporte_2013_A}, but making use of idealised simulations of
dSph galaxies orbiting in a static MW potential. The galaxies are
initialised as perfect stellar discs, embedded in spherical dark
matter subhaloes, with the two dSph stellar populations assumed to
originate from the inner and outer disc. These authors make use of the
velocity dispersion measured within $R_\mathrm{e}$ when computing
$\Gamma$, in contrast to the method of \citet{Walker_2011}, which uses
the dispersion averaged over the whole
galaxy.\footnote{\citet{Kow_2013} refer to the estimator they use as
  that of \citet{Wolf_2010}. In fact, \citet{Wolf_2010} advocate
  averaging the dispersion over all stars belonging to the galaxy.}
They find that $\Gamma$ can be biased both low \textit{and} high with
respect to the true slope, depending on the line of sight, and that
the results are most accurate for galaxies that are close to
spherical. If $\Gamma$ was overestimated for Sculptor or Fornax by
\citet{Walker_2011} then the confidence with which they exclude an
inner NFW slope would be artificially high. However, as pointed out by
\citet{Laporte_2013_A}, since \citet{Kow_2013} use the dispersion
measured within $R_\mathrm{e}$ for each population, this may yield
significantly different results to the method used by
\citet{Walker_2011}. Indeed, \citet{Laporte_2013_A} demonstrate that
if they measure the dispersion within $R_\mathrm{e}$ then their
analysis using the Aquarius simulations can also overestimate
$\Gamma$. Note that in this case $\Gamma$ is \textit{more accurate}
(i.e.\ \textit{less biased}), albeit typically with a larger scatter.
However, since the objective of applying the methodology of
\citet{Walker_2011} is to place \textit{conservative} upper limits on
the inner density profile slopes, in the context of ruling out NFW
profiles, more accurate values of $\Gamma$ are deemed to be
undesirable, since in this case the slope is much more likely to be
\textit{overestimated}, hence leading to unreliable (i.e.\
non-conservative) NFW profile exclusion confidence values.

\textit{If} it can be assumed that the biases in the estimated masses
for the two stellar populations are each determined \textit{entirely}
by the `choice' of sight-line orientation relative to the whole
dynamical system (galaxy) for a single observer, and no other factors
are relevant (including e.g.\ the 3D shapes and relative orientations
of the distinct stellar populations), then the mass bias will of
course be identical for the two stellar populations,\footnote{Ignoring
  from now on the claimed dependence on the stellar population
  concentration within the halo ($r_\mathrm{half}/r_\mathrm{DM}$),
  since this acts to make the NFW exclusion limits more conservative.}
and the line of sight choice will not affect the recovered $\Gamma$
value. However, it seems somewhat unlikely that this assumption would
hold in general for real galaxies, since the two metallicity
populations are likely to have formed by different processes and at
different times \citep{Benitez_2016}.  The two populations might have
different levels of dispersion support, velocity dispersion
anisotropy, 3D shapes, alignments relative to their host halo and to
each other, and may be in different dynamical states relative to
equilibrium; each factor potentially introducing scatter into the two
mass estimates. Differences in these important properties between the
two populations could result in the introduction of additional
uncorrelated biases into the two recovered mass values, and hence lead
to propagation of unanticipated errors into the measurement of
$\Gamma$.

Clearly, the validity of assuming that the two populations yield mass
estimates with identical (and hence irrelevant) random biases depends
inversely on the extent to which it is valid to assume that the
populations are independent. Nonetheless, if we are to exclude NFW
profiles within dSphs with multiple stellar populations, then surely
the conservative approach is to assume that the two populations are
indeed independent, with uncorrelated random mass errors. In fact, the
APOSTLE galaxies studied here exhibit a range of alignment angles and
3D shapes relative to their host haloes (see
Fig.~\ref{fig:est_summary}). Therefore it seems entirely plausible
that the dual stellar populations found in certain dSphs could have
different orientations and shapes with respect to each other (and
their host halo), depending on their physical origins.  We plan to
return to this important issue in a subsequent paper.

In the special (and perhaps unlikely) limiting case where the two
stellar populations are in perfect morphological alignment with each
other, Fig.~\ref{fig:est_angle_grid} corresponds to the prediction
from our simulations for the mean value of the estimated to true mass
ratio for the \cite{Walker_2009} estimator, as a function of viewing
angle relative to the (here assumed perfectly aligned) principal axes
of the stellar populations. Even though this mean bias would cancel
out when computing $\Gamma$, there is also a substantial scatter
around the mean expectation, for any randomly chosen line of sight, as
shown in Fig.~\ref{fig:est_angle_grid_sdev}. Thus even for perfect
alignment of the populations with each other (and a single fixed line
of sight), the fractional errors on the estimated masses for the two
populations may be significantly different, thus perturbing the
inferred value of $\Gamma$, as a result of e.g.\ differences in the 3D
shapes, levels of dispersion support, or velocity dispersion
anisotropies of the two populations. Considering the distribution of
the standard deviation values, computed directly from the pixels shown
in Fig.~\ref{fig:est_angle_grid_sdev}, the median value of the
standard deviation (evenly weighted over the spherical surface) is
$\sigma[\alpha_\mathrm{Walker}(\theta,\phi)]=0.140^{+0.051}_{-0.028}$,
where the quoted range corresponds to the $16^\text{th}$ and
$84^\text{th}$ percentiles.  Thus under the assumption of exact
alignment of the populations, this value of
$\sigma[\alpha_\mathrm{Walker}(\theta,\phi)]$ could be taken as the
scatter in the estimated masses for the individual populations, for a
single randomly chosen line of sight. Yet clearly the expected scatter
increases as the populations move away from the state of perfect
alignment assumed in this special case, and the mean biases for
different sight-line orientations relative to the individual
populations become relevant (see Fig.~\ref{fig:est_angle_grid}).

To represent the various uncertainties involved in estimating the
enclosed masses for the two stellar populations, we assume in the
following that, for each population, the estimated mass has an
associated $1\sigma$ fractional error of $\sigma_\mathrm{sys}=0.2$. If
the important properties of the two stellar populations are assumed to
be independent, then the choice of $\sigma_\mathrm{sys}=0.2$ seems
reasonable in general given our analysis of the uncertainties involved
in estimating the dynamical masses of realistic simulated galaxies in
Section~\ref{sec:app}. This representative scatter of 20 percent is
relatively high compared to the standard deviations of the $\alpha$
values for most line of sight orientations relative to the stellar
principal axes for the galaxies in our sample, although much larger
scatters are observed for many viewing angles.  Within the context of
obtaining \textit{conservative} exclusion confidences for NFW-like
inner density profile slopes, our choice of $\sigma_\mathrm{sys}$
might in fact be considered somewhat low; in which case the results
presented below constitute an underestimate of the impact of the mass
estimator errors on the reliability of the results of
\cite{Walker_2011}.  However, we are ignorant of any underlying
correlations between the properties of the dual populations in
Sculptor or Fornax that may conspire to bias the estimated enclosed
masses in a similar way for the two radii, thus reducing the error on
the inferred mass slope.

\begin{figure}
  \centering
  \includegraphics[width=\columnwidth]{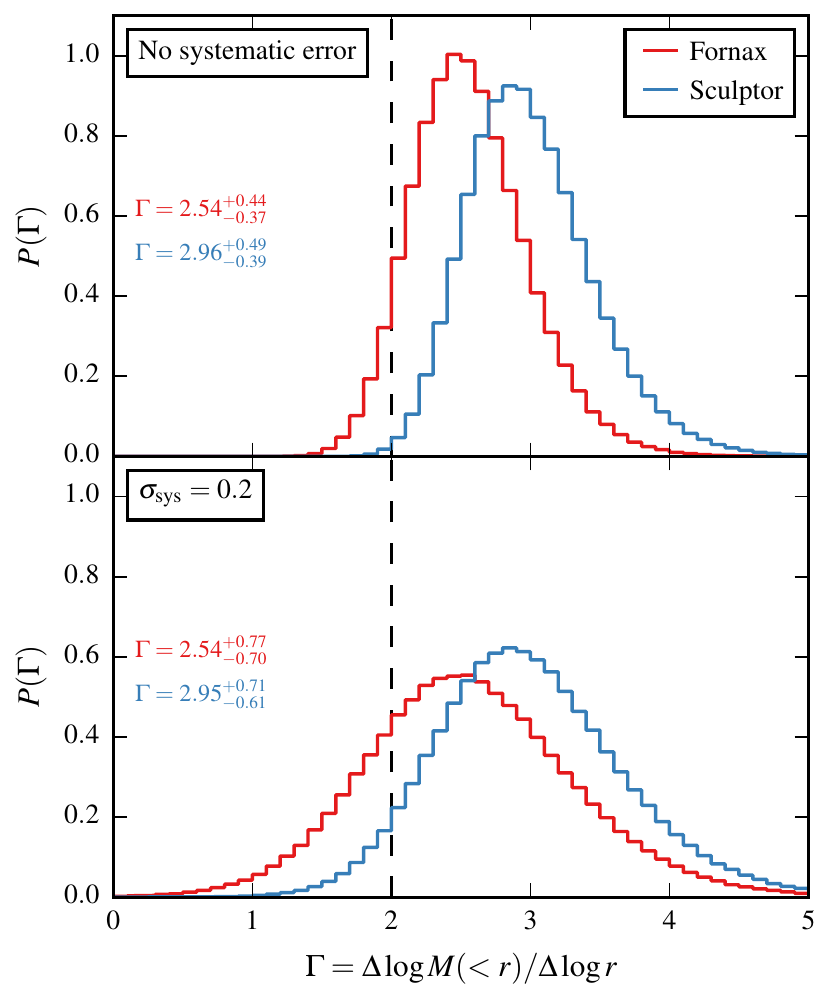}
  \caption{Distributions, $P(\Gamma)$, of the logarithmic mass profile
    slope, $\Gamma$, derived using chemo-dynamically distinct stellar
    subpopulations in Sculptor and Fornax (see
    equation~\ref{eq:slope}). The upper panel shows the distributions
    from sampling the posterior probability distribution functions for
    the half-light radii and velocity dispersions from
    \citet{Walker_2011}. The lower panel shows the results of
    repeating this procedure, but introducing a $1\sigma$ scatter of
    $\sigma_\mathrm{sys}=0.2$ in the mass estimates for both
    populations. The median $\Gamma$ value and
    $16^\text{th}-84^\text{th}$ percentile confidence interval are
    given in each panel for each distribution, in the same colours as
    the lines. Each distribution is normalised such that
    $\int_{-\infty}^\infty P(\Gamma) \mathrm{d} \Gamma =1$. The
    vertical dashed line shows the central (maximum) mass slope for an
    NFW profile.}
  \label{fig:WP11}
\end{figure}

In order to assess the impact of the systematic errors inherent in
simple mass estimators of the form assumed in
equation~(\ref{eq:slope}) on the results of \citet{Walker_2011}, we
begin by creating a simple model of their analysis
procedure. Approximating the published posterior probability
distribution functions for $\log_{10}(r_2)$, $r_1/r_2$,
$\log_{10}(\sigma_1^2)$, and $\log_{10}(\sigma_2^2)$ as Gaussian
(using the tabulated medians and $16^\text{th}-84^\text{th}$
percentile ranges), we generate $10^6$ values for $\Gamma$ by
independently drawing random values consistent with the four
distributions, and applying equation~(\ref{eq:slope}). This approach
ignores the correlations between the free parameters of the system,
which are accounted for in the MCMC chains. However, the resulting
$P(\Gamma)$ distributions for the two dSphs, shown in the upper panel
of Fig.~\ref{fig:WP11}, are very similar to those obtained by
\citet[cf.\ their fig.~10]{Walker_2011}. Indeed, our median values for
the slopes of $\Gamma=2.54^{+0.44}_{-0.37}$ for Fornax and
$\Gamma=2.96^{+0.49}_{-0.39}$ for Sculptor are in excellent agreement
with their results.

The lower panel of Fig.~\ref{fig:WP11} shows the result of including a
representative $1\sigma$ fractional systematic error of
$\sigma_\mathrm{sys}=0.2$ in the mass estimates for both
populations. In detail, we repeat the process described above, but for
each set of radii and dispersions, we multiply $M(<r_1)$ and $M(<r_2)$
(or equivalently, the squared dispersions) by random values, $E_1$ and
$E_2$, respectively, drawn separately from a Gaussian distribution
centred at one with standard deviation $\sigma_\mathrm{sys}$, and then
proceed to apply equation~(\ref{eq:slope}) as before.\footnote{We
  impose a lower limit of 0.01 for each of $E_1$ and $E_2$.} Including
the systematic scatter spreads out the $P(\Gamma)$ distributions, with
a negligible impact on the median values, giving
$\Gamma=2.54^{+0.77}_{-0.70}$ for Fornax and
$\Gamma=2.95^{+0.71}_{-0.61}$ for Sculptor. In terms of the
significance as defined in equation~(\ref{eq:signif}), the $P(\Gamma)$
distributions including the systematic error disfavour NFW slopes with
confidence of only 77.7 and 94.5 percent for Fornax and Sculptor
respectively.  Clearly, the systematic errors associated with mass
estimators of the form given in equation~(\ref{eq:est}) have the
potential to reduce the confidence with which it can be claimed that
the results of \citet{Walker_2011} rule out the presence of dark
matter cusps as steep as that of the NFW profile in Fornax and
Sculptor, depending on the extent to which the individual stellar
populations are independent from each other.

A deeper understanding of the uncertainties associated with the
methodology of \citet{Walker_2011} will come from high resolution
simulations where the chemo-dynamically distinct stellar populations
in dSphs are treated in a realistic way, ideally as the result of
cosmological initial conditions, such as in the APOSTLE simulation
suite. Dwarf galaxies with multiple stellar populations may reside in
our simulations, facilitating direct tests of methods that use
estimators of enclosed dynamical masses to infer density slopes. Such
analysis is beyond the scope of the present paper; we leave the
identification and classification of such galaxies, and analysis of
their physical origins, to future studies using APOSTLE and similar
high resolution cosmological hydrodynamical simulations.

\begin{figure}
  \centering
  \includegraphics[width=\columnwidth]{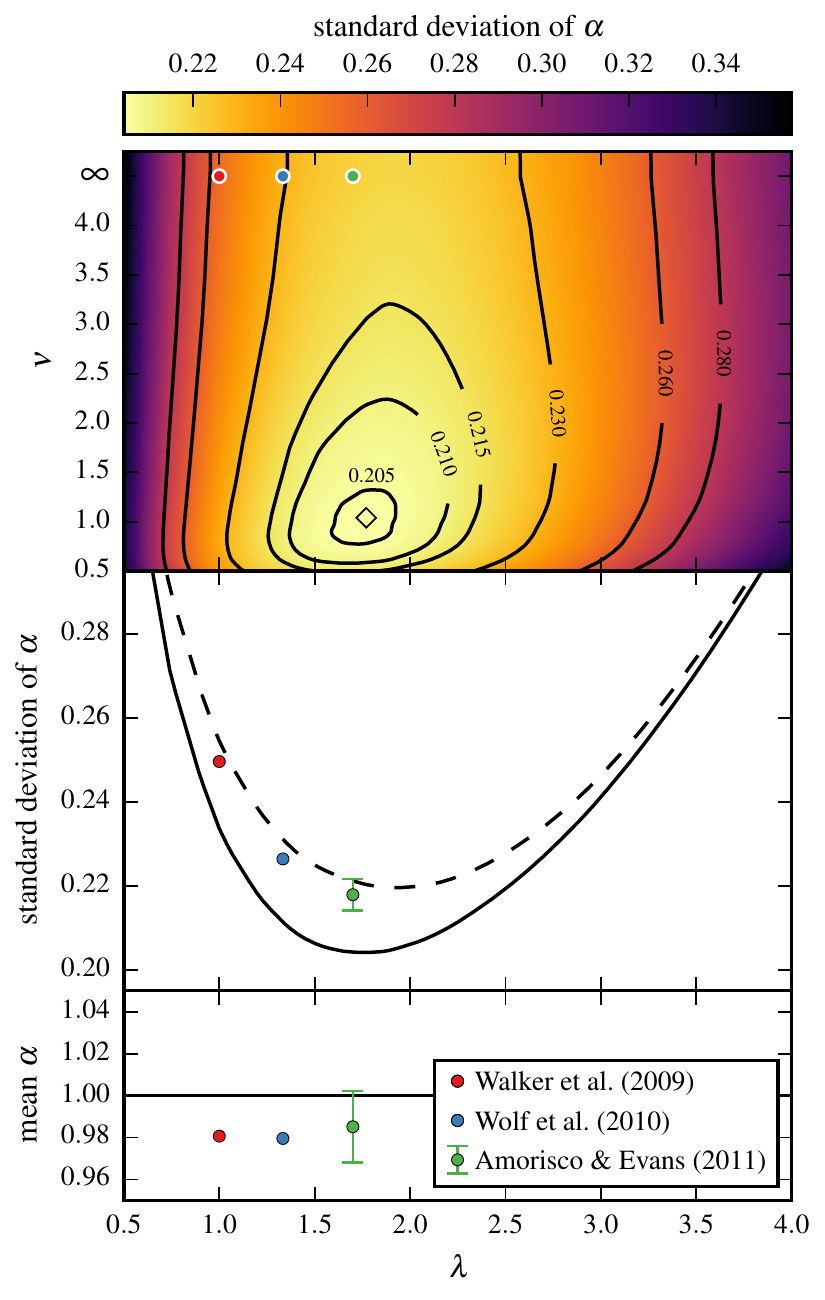}
  \caption{Calibration of mass estimator parameters ($\lambda$, $\mu$,
    $\nu$) using all dispersion-dominated galaxies in our simulated
    sample (see equation~\ref{eq:est2}). The upper panel shows the
    standard deviation of the estimated to true mass ratio $\alpha$ as
    a function of $\lambda$ and $\nu$, where the proportionality
    constant $\mu$ is chosen such that the estimator is
    \textit{unbiased} (mean $\alpha$ equals one) for each combination
    of $\lambda$ and $\nu$. The scatter shown is linearly interpolated
    between $\nu=4$ and the case where the dispersion is averaged over
    the whole galaxy ($\nu\rightarrow\infty$).  The lines show
    contours of constant standard deviation, as labelled. The diamond
    symbol indicates the location of minimum scatter (see
    Table~\ref{tab:calib}). The middle panel shows the scatter as a
    function of $\lambda$, for the choice of $\nu$ which minimizes the
    scatter at each value of $\lambda$ (solid line), and using instead
    the velocity dispersion averaged over the whole galaxy (dashed
    line). The points show the corresponding results from applying the
    estimators of \citet{Walker_2009}, \citet{Wolf_2010}, and
    \citet{Amorisco_2011}, which each use the dispersion averaged over
    the whole galaxy.  Note that the $\alpha$ distribution for a given
    estimator can have a lower scatter than the dashed line, provided
    that it is biased in the mean. The lower panel shows the mean
    $\alpha$ for each of these three estimators.}
  \label{fig:calib}
\end{figure}

\section{An optimal estimator}
\label{sec:calib}

The mass estimators discussed in the preceding sections have a common
form, as parameterized in equation~(\ref{eq:est}), where $\lambda$ is
the 3D radius within which the mass is estimated, in units of
$R_\mathrm{e}$, and $\mu$ is the estimator prefactor.  Extending this
general idea, it may be that the radius within which the velocity
dispersion is measured influences the accuracy of the recovered
mass. Including an additional dimensionless parameter, $\nu$, we can
write a more flexible generalisation,
\begin{equation}
  M(<\lambda R_\mathrm{e}) = \frac{\mu \langle\sigma_\mathrm{los}(<\nu R_\mathrm{e})\rangle^2 R_\mathrm{e}}{G}~,
  \label{eq:est2}
\end{equation}
where $\langle\sigma_\mathrm{los}(<R)\rangle$ is the line-of-sight
stellar velocity dispersion measured within a projected radius, $R$,
of the galactic centre. Up to now, we have averaged the velocity
dispersion over the whole galaxy ($\nu\rightarrow\infty$), in keeping
with the estimators proposed by \citet{Walker_2009} and
\citet{Wolf_2010}.

Setting aside the theoretical motivation for certain choices of the
parameters in equation~(\ref{eq:est2}) which, at least for the
estimators of \citet{Walker_2009} and \citet{Wolf_2010}, are based on
the assumptions that underpin the spherical Jeans equation combined
with additional simplifications, we now ask: empirically, for the
population of dispersion-dominated galaxies in our simulations, which
set of estimator parameters yields an unbiased estimate of the true
dynamical mass within some radius, with minimum scatter?

Considering our dispersion-dominated galaxy sample, the upper panel of
Fig.~\ref{fig:calib} shows the standard deviation of the estimated to
true mass ratio, $\alpha$, for the unbiased estimator obtained by
setting $\mu$ such that the mean $\alpha$ is one for a given $\lambda$
and $\nu$. This grid has been computed by projecting over \nproj\
evenly distributed lines of sight, with a grid spacing of 0.01 in
$\lambda$ and $\nu$. The scatter in the estimates varies strongly with
$\lambda$, and depends more weakly on $\nu$. However, the impact of
$\nu$ on the scatter becomes more significant near the location of
minimum scatter in this plane (diamond marker). The middle panel in
Fig.~\ref{fig:calib} shows the scatter as a function of $\lambda$ for
the case where $\nu$ is chosen to minimize the scatter for each value
of $\lambda$ (solid line), and also where the velocity dispersion
averaged over the whole galaxy is used instead (dashed line). In the
case where $\nu$ is allowed to vary, the scatter is minimized for
$\nu \approx 1$, while including the whole galaxy in the dispersion
measurement leads to a preference for higher $\lambda$.

\begin{figure}
  \centering
  \includegraphics[width=\columnwidth]{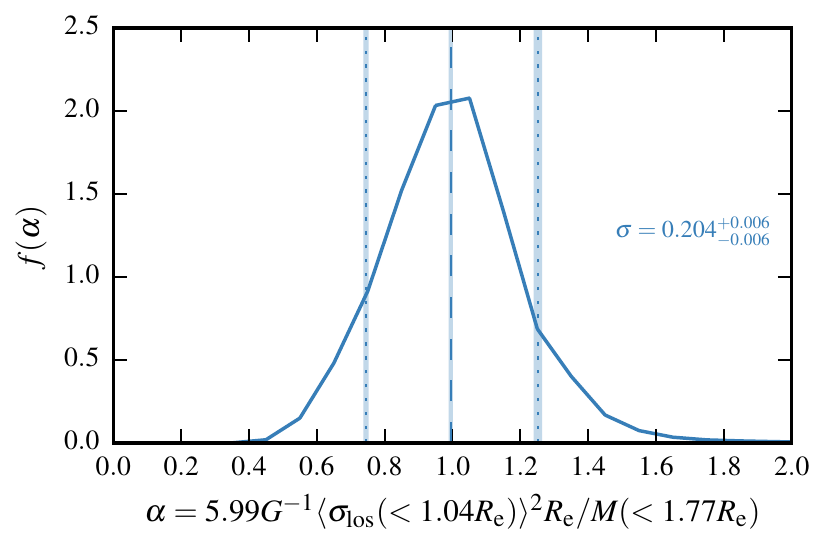}
  \caption{Distribution, $f(\alpha)$, of the estimated to true mass
    ratio, $\alpha$, for the unbiased estimator with minimum scatter
    of the form given in equation~(\ref{eq:est2}). The optimum
    parameters are given in Table~\ref{tab:calib} (as in the
    horizontal axis label). The distribution considers all
    dispersion-dominated galaxies ($\kappa_\mathrm{rot}<0.5$), using
    projections from \nproj\ evenly distributed lines of sight. The
    dashed line shows the median, and the dotted lines show the
    $10^\text{th}$ and $90^\text{th}$ percentiles. The standard
    deviation, $\sigma$, is also shown.  The shaded regions around the
    vertical lines and the quoted errors on $\sigma$ are the
    $16^\text{th}-84^\mathrm{th}$ percentile confidence limits,
    derived from applying the estimator to $10^4$ bootstrap samples of
    the galaxies.  Note that the similar $\sigma$ errors given in
    Table~\ref{tab:calib} for this estimator are instead derived from
    the distribution of optimum $\sigma$ values obtained from
    repeatedly calibrating equation~(\ref{eq:est2}) on bootstrap
    samples of the same galaxies.  Each projection of each galaxy
    contributes to the distribution with equal weight, and the
    distribution is normalised to have unit area. }
  \label{fig:calib_dist}
\end{figure}

The sets of parameters that minimize the scatter for these two
treatments of $\nu$ are given in Table~\ref{tab:calib}, along with the
associated scatter in the $\alpha$ distributions.  We shall refer to
the parameters found in the case where $\nu$ is allowed to vary as the
`optimum' set.  The parameters in the case where the dispersion is
measured within exactly $R_\mathrm{e}$ is also given in
Table~\ref{tab:calib} ($\nu=1$) and yield an equivalent level of
scatter as in the optimum case.  The uncertainties on the parameters
and the scatter values given in Table~\ref{tab:calib} are derived from
bootstrap resampling of the galaxies, where for each constraint on
$\nu$, we draw $10^4$ samples of the same number of galaxies as in the
true sample, and compute the $16^\text{th}$ and $84^\text{th}$
percentiles of the distributions of the parameters and scatter in
$\alpha$ resulting from applying the minimisation procedure to each
sample. Fig.~\ref{fig:calib_dist} shows the full $\alpha$ distribution
in the optimum case. The reduction in the minimum scatter due to
switching from $\nu\rightarrow\infty$ to the optimum parameter set is
7.0 percent.

The mean $\alpha$ values and associated scatter obtained from applying
the estimators of \citet{Walker_2009} and \citet{Wolf_2010} to the
same galaxy sample as used in the calibration are shown in the lower
panels of Fig.~\ref{fig:calib}. Both of these estimators have a
slightly smaller scatter than in the unbiased case that uses the
dispersion averaged over the whole galaxy (as these two estimators
do), for their $\lambda$. Using the optimum parameter set leads to
reductions in the scatter of 18.2 and 9.9 percent relative to the
\citet{Walker_2009} and \citet{Wolf_2010} versions, respectively,
while correcting for the bias exhibited by each estimator, as shown in
Fig.~\ref{fig:calib}.

We note that \citet{Amorisco_2011} obtain an empirical estimator with
similar parameters to our optimum result, from phase-space modelling
of the dSphs of the MW (they find $\lambda=1.7$ and $\mu=5.8\pm1.0$;
cf.\ Table~\ref{tab:calib}). The choice of $1.7R_\mathrm{e}$ for the
enclosing radius minimizes the dependence in their analysis of the
recovered mass on the assumed halo density profile and its associated
scale length, assuming that the dSphs have isotropic velocity
dispersions in their centres, with relatively flat
$\sigma_\mathrm{los}(R)$ profiles. The mean $\alpha$ and scatter
resulting from applying this estimator to our dispersion-dominated
sample are shown in Fig.~\ref{fig:calib} alongside the
\citet{Walker_2009} and \citet{Wolf_2010} results, where the error
bars indicate the spread due to the quoted uncertainty on
$\mu$.\footnote{\citet{Amorisco_2011} state that their analysis makes
  use of the velocity dispersion at the galactic centre, however it is
  not clear to what level the dispersions used can be considered to be
  `central'. The dataset used is that compiled by
  \citet{Walker_2009,Walker_2009_erratum}, who note that the
  dispersions are global averages. Therefore we use the velocity
  dispersion averaged over the whole galaxy when applying the
  estimator of \citet{Amorisco_2011}, in order to be consistent with
  the data used in their analysis.}

\begin{table}
  \centering
  \caption{Parameters ($\lambda$, $\mu$, $\nu$) that yield an
    estimated to true mass ratio, $\alpha$, of one on average, with
    minimum scatter, calibrated using all galaxies in our
    dispersion-dominated sample (see equation~\ref{eq:est2}). The
    three sets of parameters shown result from different constraints
    on $\nu$, which sets the radius within which the velocity
    dispersion is measured. The final row gives the scatter for each
    parameter set (standard deviation of $\alpha$).  The quoted
    uncertainties are the $16^\text{th}-84^\text{th}$ percentile
    confidence limits, from $10^4$ bootstrap samples of the galaxies,
    for each constraint on $\nu$.  }
  \label{tab:calib}
  \begin{tabular}{cccc}
    \hline
    Parameter & $0.1<\nu<4.0$              & $\nu=1$                    & $\nu\rightarrow\infty$ \\
              & (optimum set) \\
    \hline
    $\lambda$ & $1.77^{+0.08}_{-0.06}$     & $1.76^{+0.09}_{-0.05}$     & $1.91^{+0.09}_{-0.03}$ \\
    $\mu$     & $5.99^{+0.37}_{-0.33}$     & $5.94^{+0.42}_{-0.28}$     & $6.95^{+0.48}_{-0.18}$ \rule{0pt}{2.5ex} \\
    $\nu$     & $1.04^{+0.10}_{-0.17}$     & $1$                        & $\infty$ \rule{0pt}{2.5ex} \\
    \hline
    Scatter   & $0.204^{+0.005}_{-0.006}$  & $0.204^{+0.005}_{-0.006}$  & $0.220^{+0.005}_{-0.006}$ \\
    \hline
  \end{tabular}
\end{table}

\section{Conclusions}
\label{sec:conclusions}

In this paper we have presented the key intrinsic and observable
properties of galaxies drawn from the APOSTLE simulations of the Local
Group that are relevant to Jeans analysis in general, and in
particular to the use of simple mass estimators of the form given in
equation~(\ref{eq:est}), as advocated by \citet{Walker_2009} and
\citet{Wolf_2010}. The simulated galaxies have realistic stellar
density distributions and line-of-sight stellar velocity dispersions,
combined with a range of 3D shapes, levels of dispersion support, and
stellar velocity dispersion anisotropy.

Applying the mass estimators proposed by \citet{Walker_2009} and
\citet{Wolf_2010} to many projections of each galaxy in our sample, we
have found that the estimators each have a small bias in the estimated
to true mass ratio, $\alpha$, combined with a relatively large
scatter.  This scatter is 23 or 25 percent at the $1\sigma$ level
overall for dispersion-dominated galaxies, respectively (see
Fig.~\ref{fig:est_Krot}).

The dependence of $\alpha$ on various galaxy properties is summarised
in Fig.~\ref{fig:est_summary}. The scatter in $\alpha$ depends
strongly on the shape of the galaxy, such that the $\alpha$
distributions are sharply peaked for stellar mass distributions that
are close to spherical symmetry. However, the dependence of the
scatter on the observable 2D shape on the sky is much weaker, i.e.\ it
is difficult to identify ideal spherically symmetric systems in
projection. The scatter also increases for progressively larger radial
bias in the stellar velocity dispersion anisotropy, progressively
higher levels of rotational support, and for galaxies that exhibit
closer alignment with their host dark matter haloes.  We also find
that the scatter is lowest for galaxies that have relatively low
stellar kinetic energy for their gravitational potential energy, and
for galaxies that are highly extended, in units of their halo scale
radius. The mass estimator accuracy is more sensitive to the 3D shape
(and alignment with respect to the observer) of the galaxy than to
that of the host dark matter halo.  The dispersion-dominated
satellites of the MW and M31 analogues in the simulations yield a
smaller scatter in $\alpha$ (typically 20--22 percent) than the
dispersion-dominated field galaxies (typically 24--26 percent), as
shown in Fig.~\ref{fig:est_type}.

Considering the dependence of $\alpha$ on the orientation of the line
of sight with respect to the galaxy, we find that the dynamical masses
are overestimated with a relatively large scatter for lines of sight
coincident with the stellar morphological major axis, underestimated
with a modest scatter along the minor axis, and close to accurate with
a modest scatter for the observations along the stellar principal axis
of intermediate length (see Figs.~\ref{fig:est_angle_grid}
and~\ref{fig:est_angle_grid_sdev}).

Adopting a representative value of 20 percent for the $1\sigma$
systematic scatter in $\alpha$ for satellite galaxies, we have shown
that the systematic uncertainties inherent in the simple estimators
significantly increase the errors on the estimated dynamical masses of
dSph satellites of the MW from \citet{Wolf_2010} in cases where the
observational errors on the half-light radii and velocity dispersions
are relatively small. Including the same representative scatter within
a simple model of the analysis of \citet{Walker_2011} demonstrates
that systematic uncertainties can also greatly reduce the significance
with which the mass profile slopes derived for two independent stellar
populations in Sculptor and Fornax are inconsistent with an NFW dark
matter density profile. This result depends on the level to which the
errors on the estimated masses for the two populations can be assumed
to be independent.

Finally, we have investigated the scatter in the estimated to true
mass ratio for the set of unbiased estimators of the form given in
equation~(\ref{eq:est2}), considering all the dispersion-dominated
galaxies in our simulated sample. We find that using the optimum set
of parameters given in Table~\ref{tab:calib} results in a reduction in
the scatter of between 10 and 20 percent with respect to similar
estimators in the literature
\citep{Walker_2009,Wolf_2010,Amorisco_2011}, combined with the removal
of bias in the mean. The scatter is optimised when the line-of-sight
velocity dispersion is measured within a radius close to the projected
stellar half-mass radius, $R_\mathrm{e}$ (rather than over the whole
galaxy as is typical):
\begin{equation}
  M(< 1.77 R_\mathrm{e}) = \frac{5.99 \langle\sigma_\mathrm{los}(< 1.04 R_\mathrm{e})\rangle^2 R_\mathrm{e}}{G}~.
\end{equation}
This equation gives our optimum result in the case where the radius
within which the velocity dispersion is measured is treated as a free
parameter. If instead we fix the dispersion measurement to consider
stars within exactly $R_\mathrm{e}$, the resulting scatter is the same
as that obtained using the formally optimal parameter set (see
Table~\ref{tab:calib}).

\section*{Acknowledgements}

We would like to thank Mark Wilkinson and Manoj Kaplinghat for
interesting discussions, along with Louie Strigari, who provided the
MW dSph profiles.  We would also like to thank the referee, Matt
Walker, for helpful comments and suggestions that have improved the
paper.  This work was supported by the Science and Technology
Facilities Council (grant number ST/L00075X/1) and the European
Research Council (grant number GA~267291, `Cosmiway'). DJRC
acknowledges the support of STFC studentship ST/K501979/1. This work
used the DiRAC Data Centric system at Durham University, operated by
the Institute for Computational Cosmology on behalf of the STFC DiRAC
HPC Facility (www.dirac.ac.uk). This equipment was funded by BIS
National E-infrastructure capital grant ST/K00042X/1, STFC capital
grants ST/H008519/1 and ST/K00087X/1, STFC DiRAC Operations grant
ST/K003267/1 and Durham University. DiRAC is part of the National
E-Infrastructure.



\bibliographystyle{mnras} \bibliography{references}



\appendix

\section{Wolf~et~al.~estimator in 3D}
\label{sec:wolf3D}

In the main body of this paper, we have made use of the estimator of
\citet{Wolf_2010} expressed in terms of observable quantities: i.e.\
the projected stellar half-light (half-mass) radius, $R_\mathrm{e}$,
and line-of-sight velocity dispersion,
$\langle\sigma_\mathrm{los}\rangle$
(equation~\ref{eq:wolf2D}). However, a more fundamental version of the
estimator is given in equation~(\ref{eq:wolf3D}), from which
\citet{Wolf_2010} derive equation~(\ref{eq:wolf2D}) by simply assuming
that the 3D half-light radius, $r_\mathrm{half}$, is
$4R_\mathrm{e}/3$, which is a reasonable approximation for a range of
spherically symmetric density profiles. Yet the appropriateness of
this assumption could vary significantly for some galaxies, depending
on the shapes of their stellar density profiles. We now investigate
the change in the accuracy of the estimator of \citet{Wolf_2010}, with
respect to the standard case given in equation~(\ref{eq:wolf2D}),
when: (i) a value for $r_\mathrm{half}$ is inferred from the projected
stellar density distribution, instead of assuming
$r_\mathrm{half}=4R_\mathrm{e}/3$, and (ii) typically unobservable 3D
information on the galaxy size and velocity dispersion is used in the
mass estimation.

\subsection{Deprojection of the stellar mass profile}

In order to assess whether the assumption,
$r_\mathrm{half}=4R_\mathrm{e}/3$, is optimal for the
dispersion-dominated galaxies in our simulated sample, which have a
broad range of 3D shapes, we carry out an Abel inversion of the
projected mass distribution for all \nproj\ projections of each
galaxy. We make use of splines to model the cumulative projected
stellar mass profile, $M_\mathrm{star}(<R)$. This choice allows the
model profile to be quite general in form, in the sense that the
inference of $r_\mathrm{half}$ should not suffer from any overly
restrictive parameterization that may bias the results. In detail, we
find the quartic\footnote{So that the second derivative is both
  continuous \textit{and} smooth.} spline that interpolates exactly
through points of even spacing, $\Delta R$, on the
$M_\mathrm{star}(<R)$ profile, starting from the galactic centre. The
profile is reflected about the centre before fitting, so that the
model is well behaved in the innermost regions. Additionally, we
extend $M_\mathrm{star}(<R)$ for several $\Delta R$ beyond the
furthest star particle (where it equals the total stellar mass,
$M_\mathrm{star}$). Starting with an initial spacing of
$\Delta R = R_\mathrm{e}/2$, we recursively multiply $\Delta R$ by
0.95 and refit the spline until the following condition is satisfied,
\begin{equation}
  \frac{1}{\sigma_\mathrm{fit}^2} \sum_i \left[ M_\mathrm{model}(<R_i) - M_\mathrm{star}(<R_i) \right] ^2 < n_\mathrm{star} ~,
\end{equation}
where the sum is over all star particles in the galaxy, with projected
radii $R_i$. $M_\mathrm{model}(<R)$ is the stellar mass enclosed
within $R$ according to the spline fit, and $n_\mathrm{star}$ is the
number of star particles.  The choice of the parameter
$\sigma_\mathrm{fit}=0.01M_\mathrm{star}$ represents a compromise
between the smoothness and accuracy of the fit.

Given a model for $M_\mathrm{star}(<R)$, and assuming spherical
symmetry, the 3D stellar density profile, $\rho(r)$, follows from the
Abel integral given in equation~(\ref{eq:abel}), where,
\begin{equation}
  \frac{\mathrm{d}\Sigma(R)}{\mathrm{d}R} =
  \frac{1}{2 \pi R}
  \left[
    \frac{\mathrm{d}^2M_\mathrm{star}(<R)}{\mathrm{d}R^2} - \frac{1}{R}\frac{\mathrm{d}M_\mathrm{star}(<R)}{\mathrm{d}R}
  \right] ~.
\end{equation}
The equations can then be solved numerically for $r_\mathrm{half}$.

\begin{figure}
  \centering
  \includegraphics[width=\columnwidth]{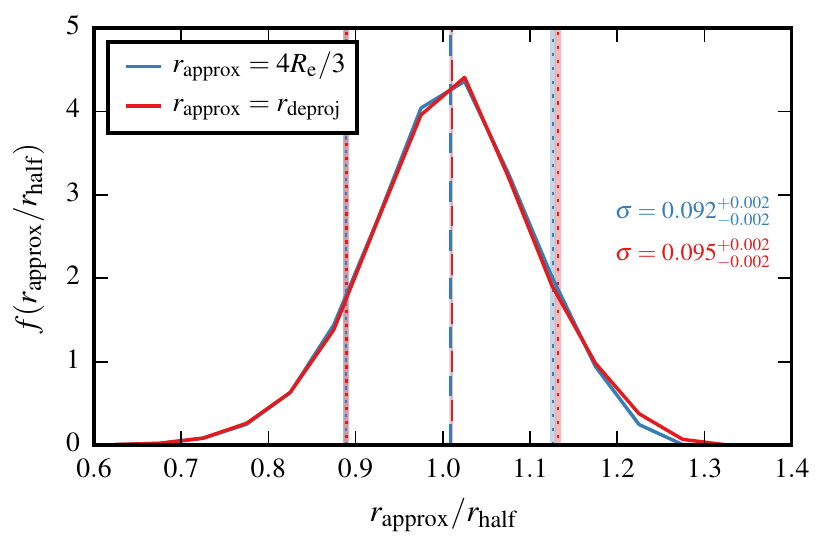}
  \caption{Distributions, $f(r_\mathrm{approx}/r_\mathrm{half})$, of
    the ratio of approximations, $r_\mathrm{approx}$, for the 3D
    half-mass radius, $r_\mathrm{half}$, to the true value of the
    radius for galaxies in our dispersion-dominated sample
    ($\kappa_\mathrm{rot}<0.5$). The $r_\mathrm{approx}$ values are
    computed for \nproj\ evenly distributed lines of sight, assuming
    either that $r_\mathrm{half}=4R_\mathrm{e}/3$, or using
    $r_\mathrm{deproj}$ from our Abel deprojection procedure
    (different colours).  Each projection of each galaxy contributes
    to the distributions with equal weight, and each distribution is
    normalised to have unit area. The vertical dashed lines show the
    median ratios, and the dotted lines show the $10^\text{th}$ and
    $90^\text{th}$ percentiles. The standard deviation, $\sigma$, of
    each distribution is given in the same colour as the lines. The
    shaded regions around the vertical lines and the $\sigma$ errors
    are the $16^\text{th}-84^\mathrm{th}$ percentile confidence
    limits, derived from $10^4$ bootstrap samples of the galaxies for
    each distribution.}
  \label{fig:rhalf}
\end{figure}

In Fig.~\ref{fig:rhalf}, we show the distribution of the ratio of the
estimated 3D half-mass radius from our deprojection procedure,
$r_\mathrm{deproj}$, to the true radius, $r_\mathrm{half}$, for all
galaxies in our dispersion-dominated sample, alongside the
corresponding distribution from assuming that $r_\mathrm{half}$ is
simply $4R_\mathrm{e}/3$.  The distributions are remarkably similar,
and show that overall there is no improvement in the approximation of
the 3D size when switching from using the factor of four thirds to
using our deprojection procedure (the scatter actually increases
slightly for $r_\mathrm{deproj}$).

\begin{figure*}
  \centering
  \includegraphics[width=\textwidth]{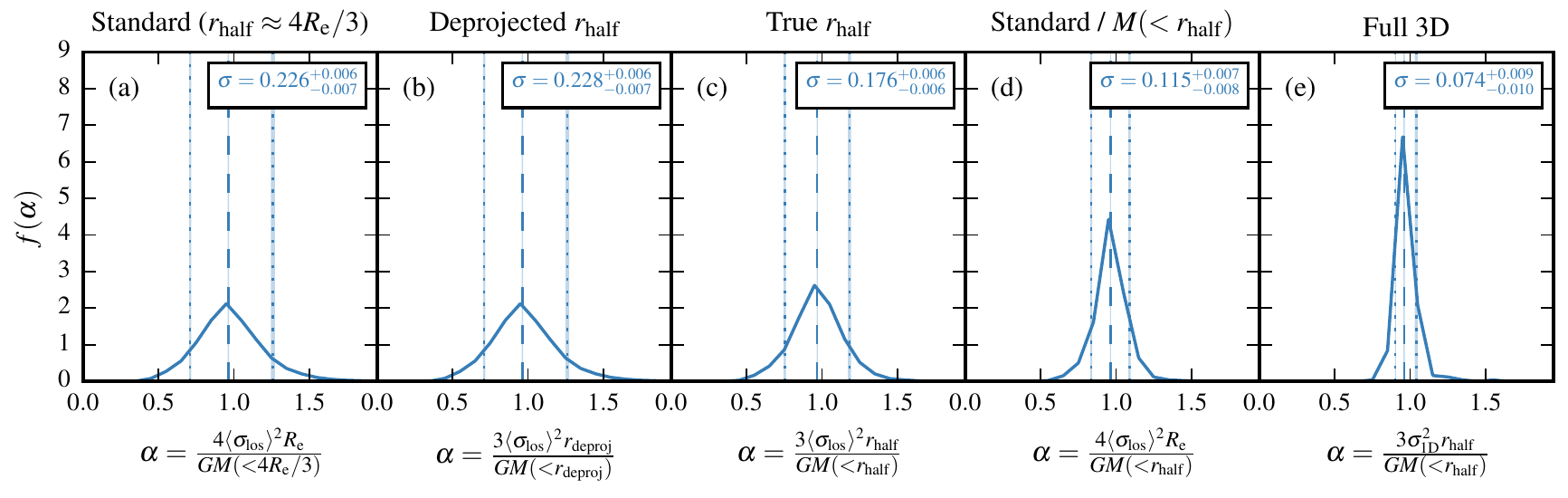}
  \caption{Distributions, $f(\alpha)$, of the estimated to true mass
    ratio, $\alpha$, for all dispersion-dominated galaxies
    ($\kappa_\mathrm{rot}<0.5$), obtained from projecting over \nproj\
    evenly distributed lines of sight, using different versions of the
    estimator proposed by \citet{Wolf_2010}. The definition of
    $\alpha$ is different in each panel, and is given explicitly in
    each horizontal axis label. The definitions are: (a) the ratio of
    the estimated mass using equation~(\ref{eq:wolf2D}) to the mass
    enclosed within the \textit{assumed} 3D half-mass radius
    ($4R_\mathrm{e}/3$) -- this is the standard definition of
    $\alpha_\mathrm{Wolf}$ as given in equation~(\ref{eq:alpha_wolf}),
    (b) the estimated to true mass ratio from
    equation~(\ref{eq:wolf3D}) using the deprojected 3D half-mass
    radius, $r_\mathrm{deproj}$, from our Abel integration procedure
    in place of $r_\mathrm{half}$, (c) the same as (b) but using the
    true 3D half-mass radius, $r_\mathrm{half}$, (d) the ratio of the
    estimated mass from equation~(\ref{eq:wolf2D}) to the mass within
    $r_\mathrm{half}$ (i.e.\ this is the ratio of the estimated to
    true mass within \textit{different spheres}, unless
    $r_\mathrm{half}=4R_\mathrm{e}/3$), and (e) the estimated to true
    mass ratio if we replace $\langle\sigma_\mathrm{los}\rangle$ in
    equation~(\ref{eq:wolf3D}) with the mean one dimensional stellar
    velocity dispersion,
    $\sigma_\mathrm{1D}=\sigma_\mathrm{3D}/\sqrt{3}$ (so no projected
    information is used in this panel).  Each projection (if
    applicable) of each galaxy contributes to the distributions with
    equal weight, and each distribution is normalised to have unit
    area. The vertical dashed lines show the median $\alpha$ values,
    and the dotted lines show the $10^\text{th}$ and $90^\text{th}$
    percentiles. The standard deviation, $\sigma$, of each
    distribution is given in each panel. The shaded regions around the
    vertical lines and the $\sigma$ errors are the
    $16^\text{th}-84^\mathrm{th}$ percentile confidence limits,
    derived from $10^4$ bootstrap samples of the galaxies for each
    distribution.  }
  \label{fig:est_wolf}
\end{figure*}

In Fig.~\ref{fig:est_wolf}(a) we show the distribution of the
estimated to true mass ratio $\alpha$ for all dispersion-dominated
galaxies in our simulated sample, obtained using the standard
projected version of the \citet{Wolf_2010} estimator from
equation~(\ref{eq:wolf2D}) (as in the upper panel of
Fig.~\ref{fig:est_Krot}). Fig.~\ref{fig:est_wolf}(b) shows the
distribution if instead we use the 3D version of the estimator from
equation~(\ref{eq:wolf3D}), replacing
$r_\mathrm{half} \rightarrow r_\mathrm{deproj}$. There is no
significant difference between the distributions shown in panels (a)
and (b), and so we conclude that assuming
$r_\mathrm{half}=4R_\mathrm{e}/3$ is already optimal, despite the
scatter shown in Fig.~\ref{fig:rhalf}.

\subsection{Using true 3D galaxy properties}
\label{sec:wolf3D_true}

It is interesting to ask how the accuracy of the estimator of
\citet{Wolf_2010} changes if intrinsic 3D, i.e.\ typically
observationally inaccessible, information is available on the galaxy
sizes and kinematics. Fig.~\ref{fig:est_wolf}(c) shows the
distribution of the estimated to true mass ratio when using the true
3D half-mass radius in equation~(\ref{eq:wolf3D}), so the only
projected quantity used to compute the distribution shown is
$\langle\sigma_\mathrm{los}\rangle$. The removal of the noise due to
the variation of the projected size reduces the scatter in the
distributions, comparing to panels (a) and (b).

In Fig.~\ref{fig:est_wolf}(d) we show the distribution of the ratio of
the estimated mass from equation~(\ref{eq:wolf2D}), i.e.\ the version
of estimator that uses $R_\mathrm{e}$, to the true mass within
$r_\mathrm{half}$. Thus, here $\alpha$ is \textit{not} the ratio of
the estimated mass to the true mass within the \textit{same} sphere,
as everywhere else in this paper.  The scatter in this distribution is
lower than in panel (c), where the only difference in the definition
of $\alpha$ is the replacement
$3 r_\mathrm{half} \rightarrow 4 R_\mathrm{e}$ from (c) to (d). That
is, equation~(\ref{eq:wolf2D}) is a better estimator of the mass
within the true 3D half-mass radius than equation~(\ref{eq:wolf3D}).
This result may seem counterintuitive at first, as the former makes
use of only projected information to estimate the mass, while the true
3D half-mass radius is used in the latter case.  This effect can be
understood in terms of the coupled variation of $R_\mathrm{e}$ and
$\langle\sigma_\mathrm{los}\rangle$ over lines of sight which view the
galaxy from many orientations, such that the variation in the
projected size counteracts that in the (squared) dispersion, so that
the product $4\langle\sigma_\mathrm{los}\rangle^2R_\mathrm{e}$ has a
smaller scatter than
$3\langle\sigma_\mathrm{los}\rangle^2r_\mathrm{half}$ for the vast
majority of galaxies (see Fig.~\ref{fig:est_angle} and also
\citealt{Laporte_2013_A}).  The result that equation~(\ref{eq:wolf2D})
is a better estimator of $M(<r_\mathrm{half})$ than
equation~(\ref{eq:wolf3D}) is only attractive if for some reason one
wanted to know $M(<r_\mathrm{half})$ without any associated estimate
of $r_\mathrm{half}$ itself!  When full 3D information is available on
the stellar velocities, so that we can replace
$\langle\sigma_\mathrm{los}\rangle$ in equation~(\ref{eq:wolf3D}) with
the mean one dimensional dispersion,
$\sigma_\mathrm{1D}=\sigma_\mathrm{3D}/\sqrt{3}$, the scatter in the
estimated masses reduces significantly, as shown in
Fig.~\ref{fig:est_wolf}(e). Note that no projected information is used
in this panel, so each galaxy contributes a single value to the
distribution.


\bsp 
\label{lastpage}
\end{document}